\documentclass[aps,pra,twocolumn,superscriptaddress,floatfix]{revtex4-1}
\usepackage{blindtext}
\usepackage{amsmath}
\DeclareMathOperator{\Tr}{Tr}
\usepackage{hyperref}
\hypersetup{
  colorlinks   = true, 
  urlcolor     = blue, 
  linkcolor    = blue, 
  citecolor   = blue 
}

\usepackage{bm}
\usepackage{gensymb}
\usepackage[normalem]{ulem}
\usepackage{euscript}
\usepackage[pdftex]{graphicx}
\usepackage{xspace}
\usepackage{xfrac}
\usepackage{enumerate}
\usepackage{braket}
\usepackage{float,fancyvrb}
\usepackage[T1]{fontenc}
\usepackage{lmodern}
\graphicspath{{./Figures/}}
\usepackage{wrapfig}
\usepackage{scrextend}
\usepackage{comment}
\usepackage{url}
\usepackage{amssymb}
\usepackage{xcolor}

\definecolor{ao}{rgb}{0.0, 0.5, 0.0}

\begin{document}

\title{Adiabatic eigenstate deformations as a sensitive probe for quantum chaos}

\author{Mohit Pandey}
\affiliation{Department of Physics, Boston University, Boston, Massachusetts, USA}
\author{Pieter W. Claeys}
\affiliation{Department of Physics, Boston University, Boston, Massachusetts, USA}
\affiliation{TCM Group, Cavendish Laboratory, University of Cambridge, Cambridge, UK}
\author{David K. Campbell}
\affiliation{Department of Physics, Boston University, Boston, Massachusetts, USA}
\author{Anatoli Polkovnikov}
\affiliation{Department of Physics, Boston University, Boston, Massachusetts, USA}
\author{Dries Sels}
\affiliation {Department of Physics, Harvard University, Cambridge MA, United States}
\affiliation{Theory of quantum and complex systems, Universiteit Antwerpen, B-2610 Antwerpen, Belgium}
\email{dsels@g.harvard.edu}
\date{\today}

\begin{abstract}
In the past decades, it was recognized that quantum chaos, which is essential for the emergence of statistical mechanics and thermodynamics, manifests itself in the effective description of the eigenstates of chaotic Hamiltonians through random matrix ensembles and the eigenstate thermalization hypothesis. Standard measures of chaos in quantum many-body systems are level statistics and the spectral form factor. In this work, we show that the norm of the adiabatic gauge potential, the generator of adiabatic deformations between eigenstates, serves as a much more sensitive measure of quantum chaos. We are able to detect transitions from non-ergodic to ergodic behavior at perturbation strengths orders of magnitude smaller than those required for standard measures. Using this alternative probe in two generic classes of spin chains, we show that the chaotic threshold decreases exponentially with system size and that one can immediately detect integrability-breaking (chaotic) perturbations by analyzing infinitesimal perturbations even at the integrable point. In some cases, small integrability-breaking is shown to lead to anomalously slow relaxation of the system, exponentially long in system size.
\end{abstract}

\maketitle

\section{Introduction}
\begin{figure}[ht]
	\centering
	\includegraphics[width= 0.48\textwidth]{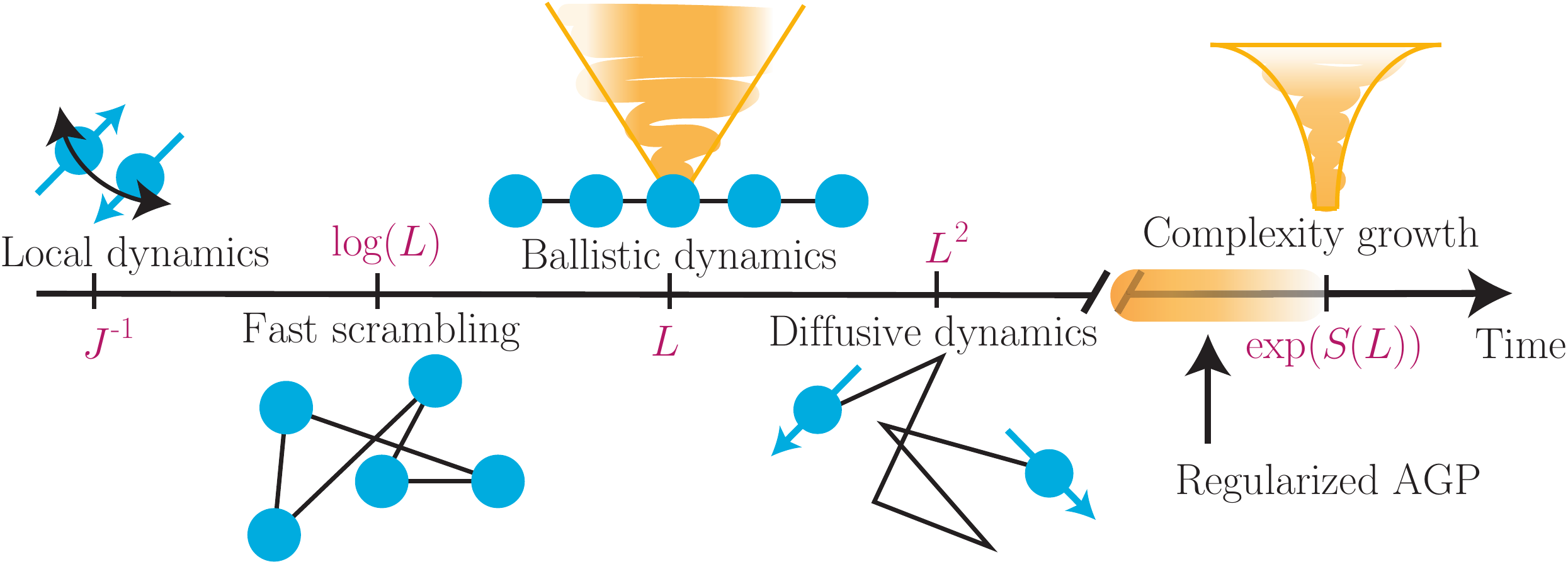}
	\caption{\textbf{Signatures of Chaos.} Quantum chaos manifests itself in a vast range of different phenomena, each relevant up to a particular system-size dependent timescale. At the earliest times, where dynamics are limited by the local bandwidth, one can see the onset of chaos. In systems without spatial locality this could lead to fast scrambling, allowing one to identify a Lyapunov exponent. Systems with spatial locality are further characterized by an additional, so-called, butterfly velocity. While this ballistic propagation ends at times $O(L)$, diffusive dynamics continues up to the Thouless time $O(L^2)$. All local dynamics has now come to a stop, nonetheless operators keep spreading over operator space, becoming increasingly more complex. This process continues for exponentially long times, only stopping at the Heisenberg time $\exp(S(L))$.}
\label{fig:intro}
\end{figure}

Finding signatures of chaos in the quantum world has been a long-standing puzzle \cite{haakequantum, stockmann1999quantum, berry1989quantum}. In the last few years exciting progress has been made on characterizing the effects of chaos on dynamical properties of quantum many-body systems, see Fig.~\ref{fig:intro} \cite{shenker_black_2014,maldacena_bound_2016,von_keyserlingk_operator_2018,rakovszky_diffusive_2018,nahum_operator_2018,swingle_unscrambling_2018,khemani_operator_2018, kudler2020conformal}. Classical chaos is usually described in terms of an exponential sensitivity of trajectories to initial conditions \cite{vulpiani2010chaos}. 
However, the quantum world precludes any definition of chaos in terms of physical trajectories due to the Heisenberg uncertainty principle. Alternatively, chaos can  be defined in terms of the absence of integrability. Classical Liouville-Arnold integrability is formulated in terms of independent Poisson-commuting integrals of motion. Again, although there have been many attempts to characterize quantum integrability in a similar way, no such unique definition exists \cite{caux2011remarks,yuzbashyan2013quantum,yuzbashyan2016rotationally,ilievski2016quasilocal}. 

In the last two decades, Random Matrix Theory (RMT) \cite{brody1981random, guhr1998random,mehta2004random} has shown outstanding success in the understanding of quantum chaos. Following the work of Wigner \cite{wigner1958distribution, wigner1993characteristic},  Bohigas, Giannoni, and Schmit \cite{bohigas1984characterization} conjectured that the energy-level statistics  of all quantum systems whose classical analogues are chaotic should show level repulsion and belong to one of three universal classes depending upon their symmetry: the Gaussian orthogonal ensemble, the Gaussian unitary ensemble, or the Gaussian symplectic ensemble. On the other hand, according to the Berry-Tabor conjecture \cite{berry1977level}, integrable systems have uncorrelated energy levels and usually exhibit Poissonian level spacing statistics. These ideas were later extended  to generic quantum systems and tested numerically under the general framework of the eigenstate thermalization hypothesis (ETH) \cite{deutsch1991quantum, srednicki1994chaos, rigol2008thermalization, borgonovi2016quantum, d2016quantum, deutsch2018eigenstate}. By now, the emergence of the random matrix behavior of quantum eigenstates is an accepted definition of quantum chaos.

Numerically, two additional steps are required before one can accurately compare the statistical properties (e.g. through level statistics or the spectral form factor \cite{muller_semiclassical_2004,bertini_exact_2018}) of a particular quantum system  with the predictions of RMT: (1) remove any symmetries; and (2) rescale the spectrum, setting the local mean level spacing to unity (also called unfolding the spectrum). Firstly, if symmetries are not removed, energy levels in different symmetry sectors don't have any correlations, so that spectra of chaotic systems can show Poissonian distributions \cite{gubin2012quantum, kudo2003unexpected}. However, finding \textit{all} symmetries of a many-body Hamiltonian is computationally hard without any physical intuition, since this effectively involves searching for all possible (local) operators that commute with the Hamiltonian. Secondly, there are various methods to unfold the spectrum, and it is known that statistics, especially ones measuring long-range correlations, can be sensitive to the adopted unfolding procedure \cite{gomez2002misleading}. Moreover, the procedure can also exhibit finite-size effects. In light of these issues, it is advisable rather to use the ratio of two consecutive level spacings \cite{oganesyan2007localization,atas2013distribution} or survival probability (see Ref. \cite{Torres-Herrera_2017,schiulaz2019thouless}).  

Here we propose an alternative tool to detect chaos in quantum systems, based on the rate of deformations of eigenstates under infinitesimal perturbations. Mathematically, the distance between nearby eigenstates (also known as the Fubini-Study metric \cite{kolodrubetz2017geometry, page1987geometrical, kobayashi1963foundations, provost1980riemannian} ) can be expressed as the Frobenius norm of the  so-called adiabatic gauge potential (AGP) \cite{berry2009transitionless,demirplak2005assisted, demirplak2003adiabatic, kolodrubetz2017geometry}, which is exactly the operator that generates such deformations. It is straightforward to show that this norm should scale exponentially with the system size in chaotic systems satisfying ETH ~\cite{kolodrubetz2017geometry}. In this sense, quantum chaos manifests itself through an exponential sensitivity of the eigenstates to infinitesimal perturbations, which can be viewed as an analogue to classical chaos, reflected in the exponential sensitivity of trajectories to such perturbations. Moreover, unlike standard probes of RMT such as the spectral form factor (see e.g. Ref.~\cite{vsuntajs2019quantum}) or the closely related survival probability (see Ref.~\cite{Torres-Herrera_2017,schiulaz2019thouless}), as well as level statistics, which only depend on the eigenvalues of the Hamiltonian, the AGP norm is sensitive to both the level spacings and the specific kind of adiabatic deformation (perturbation). 

We find that the norm of the AGP shows a remarkably different, and extremely sensitive, scaling with system size for integrable and chaotic systems: polynomial versus exponential. In our method, we do not need to remove any symmetries before computing the AGP norm needed in the analysis of the level spacing distributions and do not need to average over different Hamiltonians, which is necessary to analyze the (non self-averaging) spectral form factor. We show that one can detect chaos through the sharp crossover between the polynomial and exponential scaling of the norm. The sensitivity of this norm to chaotic perturbations is orders of magnitude greater than that of the aforementioned methods. Using this approach, we find several, previously-unexpected, results for a particular but fairly generic integrable XXZ spin chain with additional small perturbations:  i) The strength of the integrability-breaking perturbation scales exponentially down with the system size, much faster than in previous estimates \cite{modak2014finite,Modak2014}; ii)  integrability-breaking deformations immediately lead to an exponential scaling of the norm of the AGP, showing that chaotic perturbations can be already detected in the integrable regimes; and iii) in the presence of small integrability-breaking terms, the system can exhibit exponentially slow relaxation dynamics, which is similar to the slow dynamics observed in some classical nearly-integrable systems like the Fermi-Pasta-Ulam-Tsingou (FPUT) chain \cite{gallavotti2007fermi,danieli2017intermittent, pace2019behavior}. We also find that such relaxation dynamics are very different for observables conjugate (see Eq.~\eqref{AGP_orign_def} below) to integrable and chaotic directions (perturbations) of the Hamiltonian. We find similar results for an Ising model, where the integrability is broken by introducing a longitudinal field.

The connection with relaxation is not surprising, since one representation of the AGP is in terms of the long-time evolution of a local operator conjugate to the coupling. Hence, our results relate to recent studies of information propagation through operator growth in quantum many-body systems \cite{eisert2015quantum, swingle2018unscrambling,lewis2019dynamics}, where chaotic and integrable systems are again expected to exhibit qualitatively different behavior (e.g. in operator entanglement \cite{zhou2017operator,PhysRevLett.122.250603} and Lanczos coefficients \cite{parker2018universal,avdoshkin2019euclidean}). Whereas most of the previous works focused mainly on short-time effects, here we effectively focus on dynamics and operator growth at times that are exponentially long in the system size (Fig.~\ref{fig:intro}).

\section{Adiabatic Gauge Potential}
Before proceeding, let us define the adiabatic gauge potential (AGP) and discuss some of its key properties. Given a Hamiltonian $H(\lambda)$ depending on a parameter $\lambda$, the adiabatic evolution of its eigenstates as we vary this parameter is generated by the AGP as (in units with $\hbar=1$):
\begin{equation}
   \mathcal A_\lambda |n(\lambda)\rangle = i \partial_\lambda |n(\lambda)\rangle, \quad H(\lambda) |n(\lambda)\rangle=E_n(\lambda) |n(\lambda)\rangle.
\end{equation}

Using the Hellmann-Feynman theorem, it is easy to see that the matrix elements of the AGP between such eigenstates are given by 
\begin{equation}
\langle m |\mathcal{A_{\lambda}} |  n \rangle = -\frac{i}{\omega_{mn}}\langle m  |\partial_{\lambda}H | n  \rangle,
\label{AGP_orign_def}
\end{equation}
where $\omega_{mn}= E_m(\lambda)-E_n(\lambda)$, $\partial_{\lambda}H $ is the operator conjugate to the coupling $\lambda$, and we have made the dependence on $\lambda$ implicit. The diagonal elements of $\mathcal A_\lambda$ can be chosen arbitrarily due to the gauge freedom in defining the phases of eigenstates. A convenient choice consists of setting all diagonal elements equal to zero. For simplicity we will assume there are no degeneracies in the spectrum, but as will be clear shortly, this assumption is not necessary and does not affect any of the results below. We define the $L_2$ (Frobenius) norm, also called Hilbert--Schmidt norm,  of this operator as:
\begin{equation}
    ||\mathcal A_\lambda||^2=\dfrac{1}  {\mathcal{D}} \sum_n \sum_{m\neq n} |\langle n | \mathcal A_\lambda |m\rangle|^2,
    \label{eq:L2_norm}
\end{equation}
where $\mathcal{D}$ is the dimension of the Hilbert space. 

This expression should scale exponentially with the system size in chaotic systems satisfying ETH: $||\mathcal A_\lambda||^2\sim \exp\left[S\right]$, where $S$ is the entropy of the system \cite{kolodrubetz2017geometry}. Within ETH, the off-diagonal matrix elements of local operators, including $\partial_{\lambda}H$, scale as $\langle m  |\partial_{\lambda}H | n \rangle\propto \exp[-S/2]$ \cite{srednicki1994chaos,d2016quantum} while the minimum energy gap between states, $\omega_{mn}$, scales as $\exp\left[-S\right]$. The scaling of individual matrix elements was already explored in the literature to study the crossover between chaotic and non-ergodic behavior, e.g. in the context of disordered systems \cite{serbyn2015criterion,crowley2019avalanche}. 
As we will demonstrate, the exponential scaling of the norm of the AGP can be used to detect the emergence of chaotic behavior in the system with tremendous (exponential) precision.

However, Eq. ~\eqref{AGP_orign_def} is not particularly convenient: the norm of the exact AGP can be dominated by the smallest energy difference between eigenstates, and as such it is highly unstable and difficult to analyze, especially close to the ergodicity transition. Accidental degeneracies in the spectrum that are lifted by $\partial_{\lambda} H$ also cause the norm to formally be infinite. To resolve this issue, it is convenient to instead define a `regularized' AGP as follows:
\begin{eqnarray}
\langle m |\mathcal{A_{\lambda}}(\mu) |  n \rangle=  -i  \dfrac{\omega_{mn}}{ \omega_{mn}^2 + \mu^2} \langle m | \partial_{\lambda}H  | n \rangle ,
\label{AGP_mu_def}
\end{eqnarray}
where $\mu$ is a small energy cutoff. For the sake of brevity, we are going to drop the argument $\mu$ and unless specified otherwise $\mathcal{A}_\lambda$ refers to the regularized AGP. This has a clear physical intuition: instead of considering transitions (matrix elements) between individual eigenstates, we now only consider transitions between energy shells with width $\mu$. For eigenstates with $|\omega_{mn}| \gg \mu$, this reproduces the exact AGP, whereas in the limit $|\omega_{mn}| \ll \mu$, the AGP no longer diverges but reduces to a constant. Alternatively, within the operator growth representation (see Eq.~\eqref{AGP_operator_spreading} below), $\mu^{-1}$ has the interpretation of a cutoff time. Numerically, this regularization has the immediate advantage that it gets rid of any problem with (near-)divergences. Note that $\mu$ does not need to be system-size independent for this. Interestingly, as long as $\mu\propto \exp[-S]$, the norm of the AGP within chaotic systems should also remain proportional to $\exp[S]$. We can use this flexibility in defining $\mu$ to our advantage, choosing it to be parametrically larger than the level spacing to eliminate any effect of accidental degeneracies, but still exponentially small to minimize the deviation from the exact AGP. We find that choosing $\mu(L) \propto L \exp[-S(L)]$, where $L$ is the system size, is the most convenient choice (see Appendix \ref{append.muScaling}).

From Eqs.~\eqref{eq:L2_norm} and \eqref{AGP_mu_def} the norm of the regularized AGP reads
\begin{align}
    ||\mathcal A_\lambda ||^2&={1\over \mathcal D}\sum_n \sum_{m\neq n} {\omega_{mn}^2\over (\omega_{nm}^2+\mu^2)^2} |\langle m|\partial_\lambda H |n\rangle|^2\\
   & =\int_{-\infty}^\infty d\omega \dfrac{\omega^2}{  (\omega^2+\mu^2)^2}\overline{|f_\lambda(\omega)|^2},
    \label{eq:norm_agp_f_omega}
\end{align} 
where in the second equation we replaced the summation with an integration over the energy difference $\omega_{mn}=E_m(\lambda)-E_n(\lambda)$ and also defined the response function
\begin{eqnarray}
\overline{|f_\lambda(\omega)|^2}&=&{1\over \mathcal D} \sum_n \sum_{m\neq n} |\langle n | \partial_\lambda H| m\rangle|^2 \delta(\omega_{nm}-\omega) \\
&=& \dfrac{1}{\mathcal{D}} \sum_n \int_{-\infty}^{\infty} \frac{dt}{4 \pi}\, e^{i \omega t}  \langle n|\{\partial_{\lambda} H (t), \partial_{\lambda} H(0)\}| n \rangle_c, \nonumber
\label{eq:f_omega_def}
\end{eqnarray}
where $\{...\}$ stands for the anti-commutator and connected correlation function $\langle n|\partial_{\lambda} H (t) \partial_{\lambda} H(0)| n \rangle_c=\langle n|\partial_{\lambda} H (t) \partial_{\lambda} H(0)| n \rangle-\langle n|\partial_{\lambda} H (t)|n \rangle \langle n |\partial_{\lambda} H(0)| n \rangle$. Formally, this function represents an average over eigenstates $n$ of the sum of the squares of the off-diagonal matrix elements $|\langle n | \partial_\lambda H| m\rangle|^2$ with a fixed energy difference $\omega_{mn}=\omega$, which can also be obtained as the Fourier transform of the non-equal time correlation function of $\partial_{\lambda}H$.
Within the ETH ansatz, this function exactly coincides with the (averaged over eigenstates) square of the function $f_\lambda(\omega)$ introduced by M. Srednicki \cite{srednicki1994chaos}, 
according to
\begin{align}
&\langle m| \partial_\lambda H |n\rangle= f_\lambda(\omega, \bar{E}) \mathrm e^{-S (\bar E)/2} \sigma_{mn}, \label{eq:fwETH}\\
&\qquad\omega=E_m-E_n,\; \bar E=(E_n+E_m)/2,
\end{align}
with $\sigma_{nm}$ a random variable with zero mean and unit variance. Recently it was shown that the function   $|f_\lambda(\omega)|^2$ remains well defined and smooth in generic integrable systems~\cite{leblond2019entanglement,brenes_2020,brenes_2020a}.

Alternatively, it is convenient to rewrite the regularized AGP as a time integral~\cite{berry1993mv,jarzynski1995geometric, claeys2019floquet}:
\begin{equation}
\mathcal{A}_{\lambda} = - \dfrac{1}{2}  \int_{-\infty}^{\infty} dt \mathrm \, {\rm sgn}(t)\,e^{-\mu |t|}\,
\left(\partial_{\lambda}H\right) (t),
\label{AGP_operator_spreading}
\end{equation}
where ${\rm sgn}(t)$ is the sign function and
\begin{equation}
(\partial_{\lambda}H) (t)= \mathrm e^{i H t} (\partial_{\lambda}H) \mathrm e^{-i H t}
\end{equation}
is the operator conjugate to the coupling $\lambda$ in the Heisenberg representation. The exponential factor $\exp[-\mu |t|]$ can be seen as a particular choice of a filter function in the context of quasi-adiabatic continuation~\cite{hastings2010quasi,Nachtergaele2012,deroeck2017}. Notably, Eq.~\eqref{AGP_operator_spreading} remains valid for classical systems ~\cite{berry1993mv, jarzynski1995geometric} and therefore the scaling of the AGP norm can be used to detect classical chaos, which we leave for future work.

Further, Eq.~\eqref{AGP_operator_spreading}, makes clear that the inverse of the parameter $\mu$ plays the role of a cutoff time, limiting the growth of $(\partial_\lambda H) (t)$ in the operator space. Note that this time is much longer than the time scales generally studied in literature (e.g, the time scale characterizing the ballistic propagation of information $t_{LR}=L/v_{LR}$, where $v_{LR}$ is the Lieb-Robinson velocity  and $L$ is the system size)\cite{eisert2015quantum, swingle2018unscrambling,lewis2019dynamics}.  One of the outcomes of our work is that an exponential sensitivity to detecting the onset of chaos requires access to exponentially long time scales  (Fig.~\ref{fig:intro}).

\section{Numerical results}
We can now compare with results for the AGP in integrable/non-ergodic models. Specifically, we move to the analysis of the norm of the regularized AGP for a specific integrable XXZ model with open boundary conditions \cite{orbach_linear_1958,yang_one-dimensional_1966,sutherland_beautiful_2004,gaudin_bethe_2014,franchini2017introduction}, whose Hamiltonian is given below: 
 \begin{equation}
     H_{\text{XXZ}}=\sum_{i=1}^{L-1} ( \sigma_{i+1}^x \sigma_{i}^x + \sigma_{i+1}^y \sigma_{i}^y) + \Delta \sum_{i=1}^{L-1} \sigma_{i+1}^z \sigma_{i}^z. 
\label{Ham.XXZ}
 \end{equation}
We will now consider the effects of various integrability-breaking terms. Although the thermodynamics of the above model can be solved exactly using the Bethe ansatz \cite{orbach_linear_1958,yang_one-dimensional_1966,sutherland_beautiful_2004,gaudin_bethe_2014,franchini2017introduction}, we still don't have access to matrix elements of general local operators $\langle n | \partial_{\lambda}H  | m \rangle$ and the exact AGP remains out of reach even in the integrable limit. Consequently, there are also no results on the scaling of the AGP with increasing system size.
 
 For reference, we also analyze an Ising model in the presence of a longitudinal field whose Hamiltonian is given below: 
 \begin{equation}
     H_{\text{Ising}}=\sum_{i=1}^{L-1} \sigma_{i+1}^z \sigma_{i}^z + h_z \sum_{i=1}^L \sigma^z_i +h_x \sum_{i=1}^L  \sigma_i^x.
     \label{Ham.chaoticIsing}
 \end{equation}
 where open boundary conditions are chosen for the chaotic Ising model.  This model has a trivially-integrable limit at zero longitudinal field $h_z=0$, which maps to a system of free fermions \cite{sachdev2007quantum}. In this non-interacting (free) limit, the AGP can be computed analytically  \cite{del2012assisted, kolodrubetz2017geometry} (see Appendix \ref{append.free}). In the presence of the longitudinal field, this model shows a Wigner-Dyson type distribution of the energy level spacings, which is particularly pronounced at the parameters $h_x=(\sqrt{5}+5)/8$ and $h_z=(\sqrt{5}+1)/4$ \cite{kim2013ballistic}. We will use these values when computing the AGP in the chaotic regime.

\begin{figure}[ht]
	\centering
	\includegraphics[width= 0.48\textwidth]{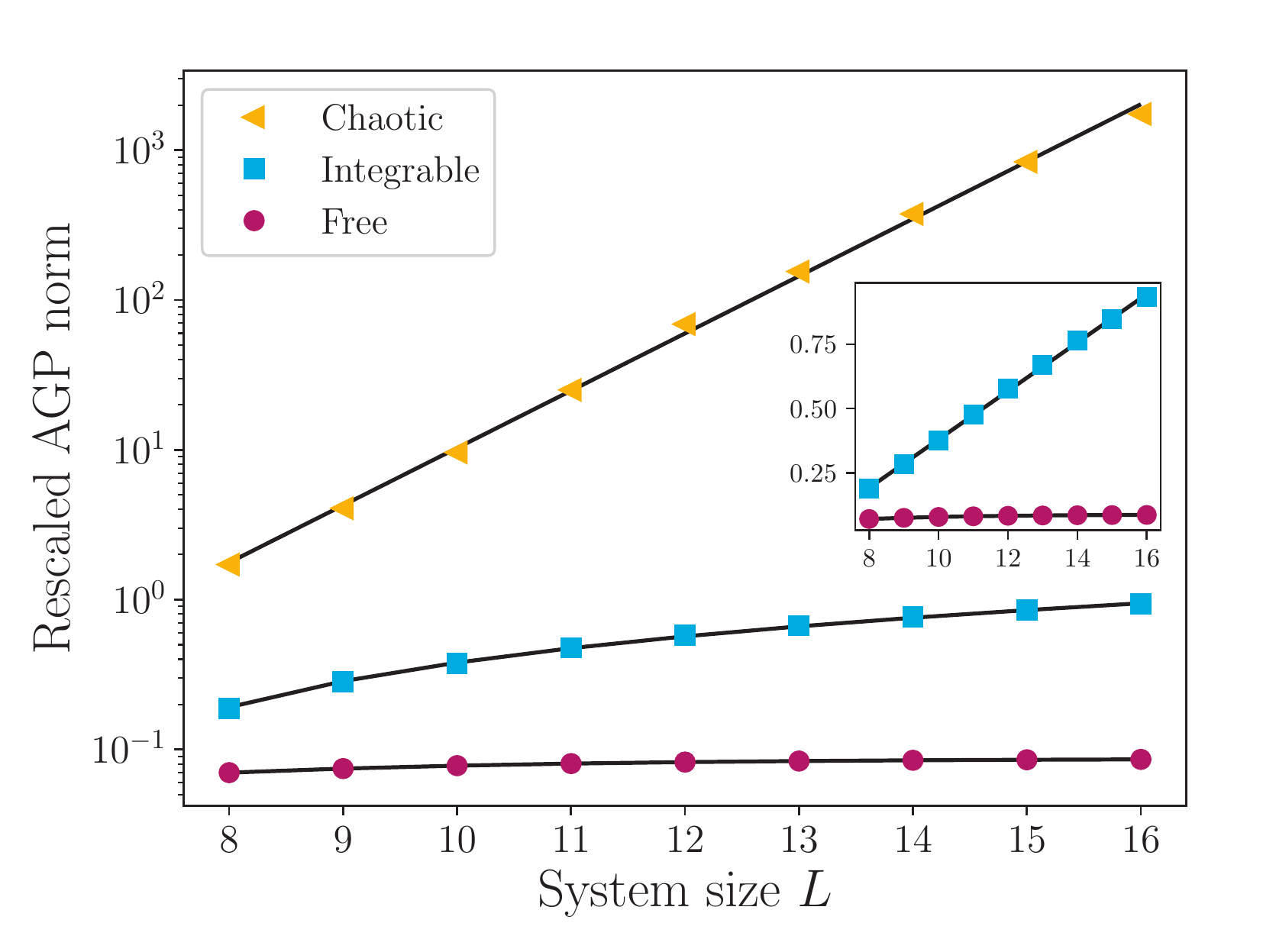}
	\caption{\textbf{AGP scaling.} The rescaled norm $||\mathcal{A}_{\lambda}||^2/L$ is presented as a function of system size for the chaotic Ising model (yellow triangles), the integrable interacting XXZ model (blue squares) and the integrable non-interacting Ising model (red dots). The data corresponding to the chaotic Ising and integrable XXZ models are fitted to an exponential and a linear function respectively (black lines). For the Ising models we set $\lambda=h_x$ and for the XXZ model we set $\lambda=\Delta$.
\textit{Inset:} Rescaled  AGP norm for the free and interacting integrable models on a linear graph. \textit{Parameters:}
$h_x=0.8$ for free model, $\Delta=1.1 $ for integrable model, $h_x=(\sqrt{5}+5)/8$ and $h_z=(\sqrt{5}+1)/4$ for chaotic model.  $||\mathcal{A}_{\text{chaotic}}||^2 \sim e^{0.9 L}$, $||\mathcal{A}_{\text{int}}||^2 =0.09 L -0.56$. }
\label{exact_regim_norm}
\end{figure}

In Fig. \ref{exact_regim_norm}, we show the AGP norm scaled by the system size $||\mathcal A_\lambda||^2/L$~\footnote{We divide the norm of the AGP by the system size for extensive perturbations, to account for the trivial extensivity of the AGP} for the interacting XXZ model and the Ising model both at the chaotic and non-interacting points. Fig. \ref{exact_regim_norm} clearly shows the remarkably different scalings with system size $L$ for chaotic, integrable and free models. For chaotic models, the scaled AGP norm shows the exponential scaling expected from ETH. For the free model, the scaled norm is system-size independent up to exponentially small corrections away from the critical point (see Appendix \ref{append.free}). For the integrable XXZ model, the scaled AGP norm shows a nontrivial polynomial scaling: $||\mathcal A_{\lambda}||^2/L\propto L^\beta$. We find that the exponent $\beta$ is non-universal and depends on the choice of the anisotropy $\Delta$ (see Appendix \ref{append.XXZ}). We have chosen $\lambda=h_x$ for both the integrable and non-integrable Ising models and $\lambda=\Delta$ for the XXZ model. 

While the exponential scaling of the AGP norm in the chaotic regime and the constant AGP norm in the free model are expected, the polynomial scaling of this norm of the XXZ integrable model is very interesting and leads to non-trivial conclusions. Recently, LeBlond {\em et al.}~\cite{leblond2019entanglement} have shown that the matrix elements of local operators in this integrable model are not sparse (as compared to the matrix elements of non-interacting integrable models). The latter implies that Eq.~\eqref{eq:norm_agp_f_omega} for the AGP norm still applies, where $|f_\lambda(\omega)|^2$ can also be found from the Fourier transform of the symmetric correlation function (see Appendix \ref{append.chaotic}). Since we have chosen $\mu$ to be exponentially small in the system size and $||\mathcal A_\lambda||^2$ is polynomially (not exponentially) large, the function $f_\lambda(\omega)$ must vanish as $\omega\to 0$. This behavior is to be contrasted with chaotic systems where at small $\omega$ this function saturates at a constant value, in agreement with the Random Matrix Theory \cite{d2016quantum}. 


\section{Integrability breaking}
Having established the scaling of the AGP norm in three different regimes, we will move to the analysis of integrability breaking by small perturbations and focus on a more generic XXZ model. As an integrability-breaking term, we choose a magnetic field coupled to a single spin in the middle of the chain, acting as a single-site defect,
\begin{equation}
    V=\sigma^z_{{\lceil}(L+1)/2 {\rceil}},
\end{equation}
where ${\lceil}(L+1)/2 {\rceil}$ stands for the smallest integer greater than or equal to $(L+1)/2$. Then we analyze the AGP for the total Hamiltonian
\begin{equation}
    H=H_{\rm XXZ}+\epsilon_d V,
    \label{Ham_XXZ_defect}
\end{equation}
as a function of the integrability-breaking parameter $\epsilon_d$. Interestingly, in Ref.~\cite{Santos2004} it was argued based on the same model that even a single site defect is sufficient to induce chaos in the thermodynamic limit. In Appendix \ref{append.NNN}, we analyze an extensive integrability-breaking  perturbation  by considering $H=H_{\rm XXZ}+\Delta_2 V$ with $V=\sum_i \sigma^z_{i+2}\sigma^z_i$ and find the results to be consistent.  The similarity between the effects of local and global perturbations on spectral properties was also found in Ref.~\cite{torres2014local}.  

\begin{figure}[ht]
	\centering
\includegraphics[width= 0.48\textwidth]{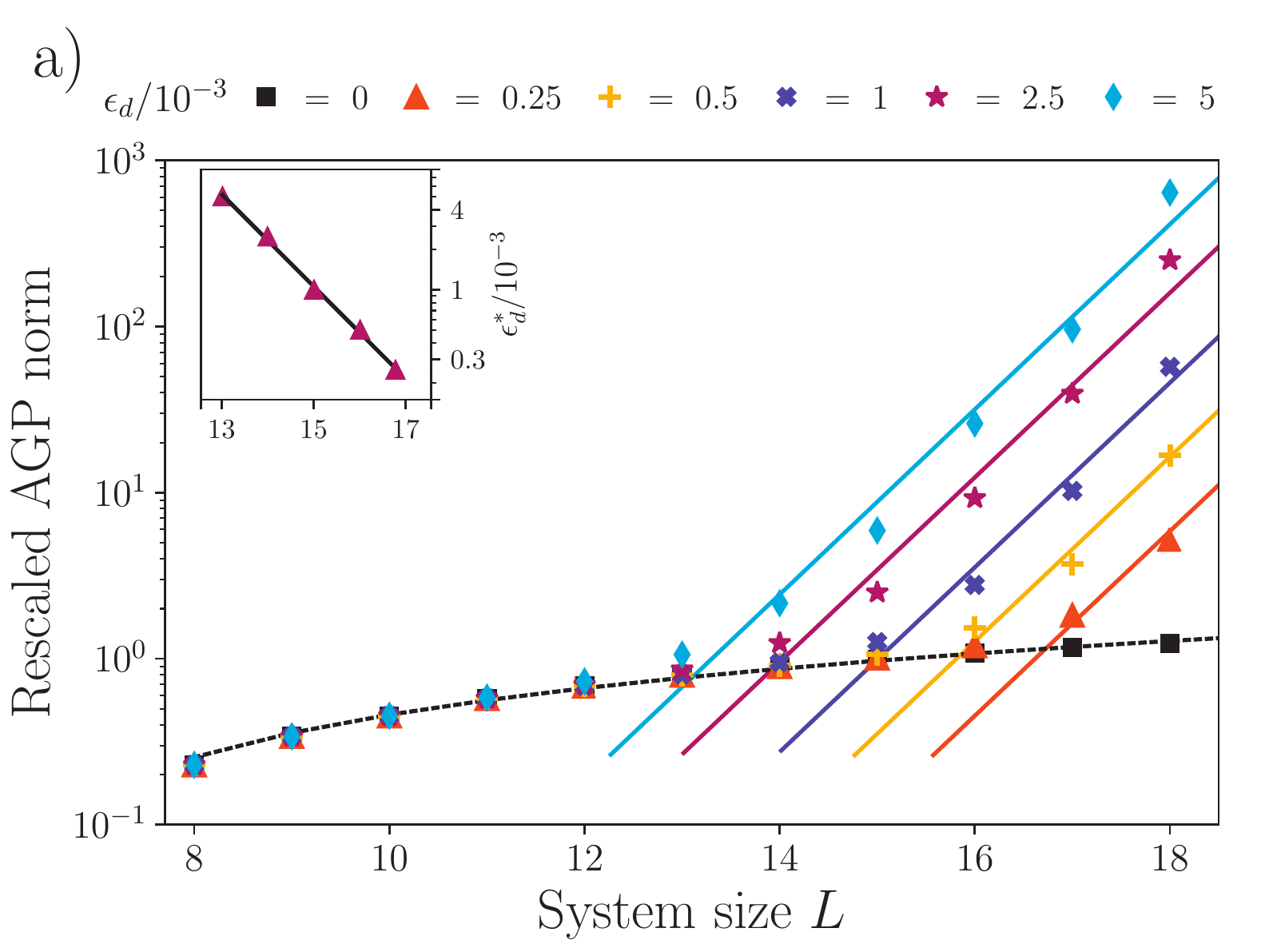}\\
\includegraphics[width= 0.48\textwidth]{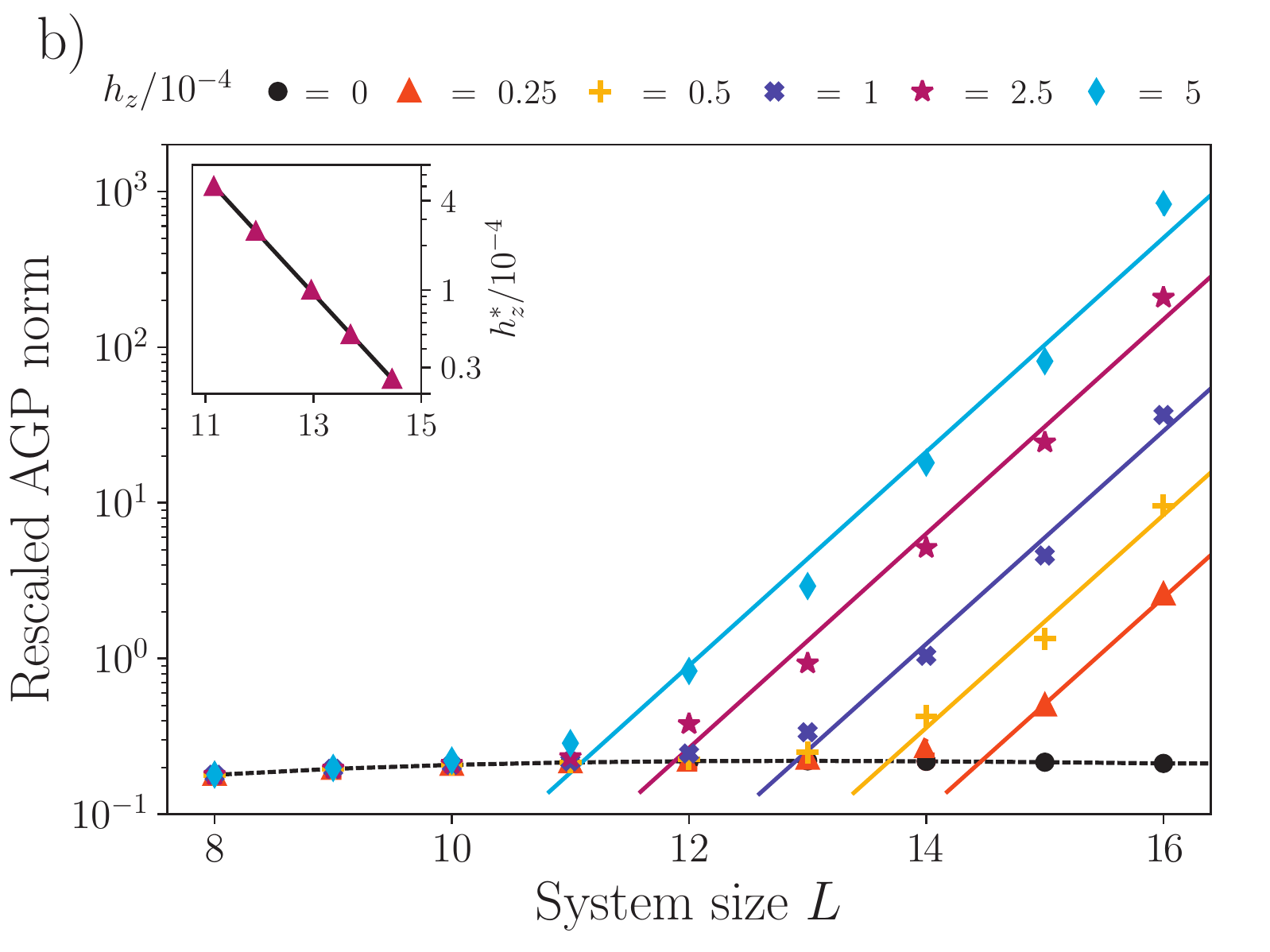}
\caption{\textbf{Integrability breaking.} The rescaled AGP norm $||\mathcal{A_\lambda}||^2/L$ of a) the XXZ chain with $\lambda=\Delta$ and b) the Ising chain with $\lambda=h_x$. Both models show a sharp crossover from polynomial to exponential scaling with system size, even for very small integrability breaking perturbation strengths. With decreasing perturbation strength, the system size where this crossover happens increases. Straight lines are the exponential fits with $||A_{\lambda}||^2/L \sim e^{ \beta L}$, where $\beta=1.28$ for the XXZ and $\beta=1.58$ for the Ising model. The insets show the scaling of the crossover point, i.e. the dependence of the integrability-breaking perturbation on system size. The critical perturbation strength scales exponentially with system size, with $\epsilon_d^* \sim e^{-0.8 L}$ for the XXZ chain and $h_z^* \sim e^{-0.9 L}$ for the Ising chain. \textit{Parameters:} a) $\Delta=1.1$, b) $h_x=0.75$.}
\label{defect_XXZ}
\end{figure}

 A challenging question is how quickly chaos emerges when a non-ergodic, or integrable system, is subjected to an integrability-breaking perturbation. In classical systems with few degrees of freedom, it is known from KAM theory that integrable systems are stable against small perturbations \cite{moser_invariant_1962,kolmogorov_conservation_1954,arnold_proof_1963}. It is widely believed that quantum chaos is generally induced by infinitesimal perturbations in the thermodynamic limit \cite{Rabson2004,Santos2010, modak2014finite,Modak2014}, with the potential exception of many-body localization \cite{nandkishore2015many, abanin2017recent}, although the precise scaling of the critical perturbation strength with the system size remains an open question. A standard limitation of numerical approaches (using e.g., level statistics or spectral form factor) addressing this question is the small system sizes amenable to simulations, where it is possible to reliably extract the data. 

In Fig.~\ref{defect_XXZ}~a) we show the scaling of the norm of the AGP as a function of the system size for different perturbation strengths $\epsilon_d$. We choose the zero magnetization subspace of the XXZ chain with number of spins up $N_{\uparrow}={\lfloor}L/2{\rfloor}$, where ${\lfloor}L/2{\rfloor}$ stands for the largest integer less than or equal to $L/2$, and for the direction of the AGP we choose $\lambda=\Delta$, i.e. as in Fig.~\ref{exact_regim_norm}. For the cutoff, we choose $\mu=L\mathcal{D}_0^{-1}$, where $\mathcal{D}_0$ is the dimension of zero magnetization sector. From the figure, we clearly see a sharp crossover in the scaling of the norm of the AGP as a function of system size from the integrable power law behavior to the chaotic exponential behavior. The straight lines are obtained by a least squares fit, with the slope extracted for the largest $\epsilon_d$ and then used for other perturbations. After the best fitting parameters were found, the critical system sizes were obtained for a particular defect energy at which the integrable (polynomial) and chaotic (exponential) curves intersect. These values are shown in the inset of Fig.~ \ref{defect_XXZ}~a), showing a clear exponential scaling of the critical perturbation strength with the system size. Interestingly, the slope of the exponential scaling $\beta\approx 1.28$ is almost twice the slope predicted by ETH, $\beta=\log(2)\approx 0.69$. Notably, the slope of $2\log(2)$ is the largest possible growth rate of the AGP norm (see Appendix \ref{append.chaotic}). In the next section we will return to this point and relate it to the emergence of relaxation times that are exponentially long in system size.

Consistent results are obtained for the Ising model~\eqref{Ham.chaoticIsing}, where one can consider breaking the integrability of the transverse field Ising model ($h_z=0$) by introducing a small non-zero $h_z$-field, while probing the integrable direction $\lambda = h_x$. The results are shown in Fig.~\ref{defect_XXZ}~b). As in the XXZ case, we observe a sharp crossover from the the unperturbed scaling of the AGP norm (see~Fig.~\ref{exact_regim_norm}) to exponential scaling with an exponent that exceeds the ETH expectation, once again having implications on the long time relaxation of the system.

\begin{figure}[ht]
	\centering
\includegraphics[width= 0.48\textwidth]{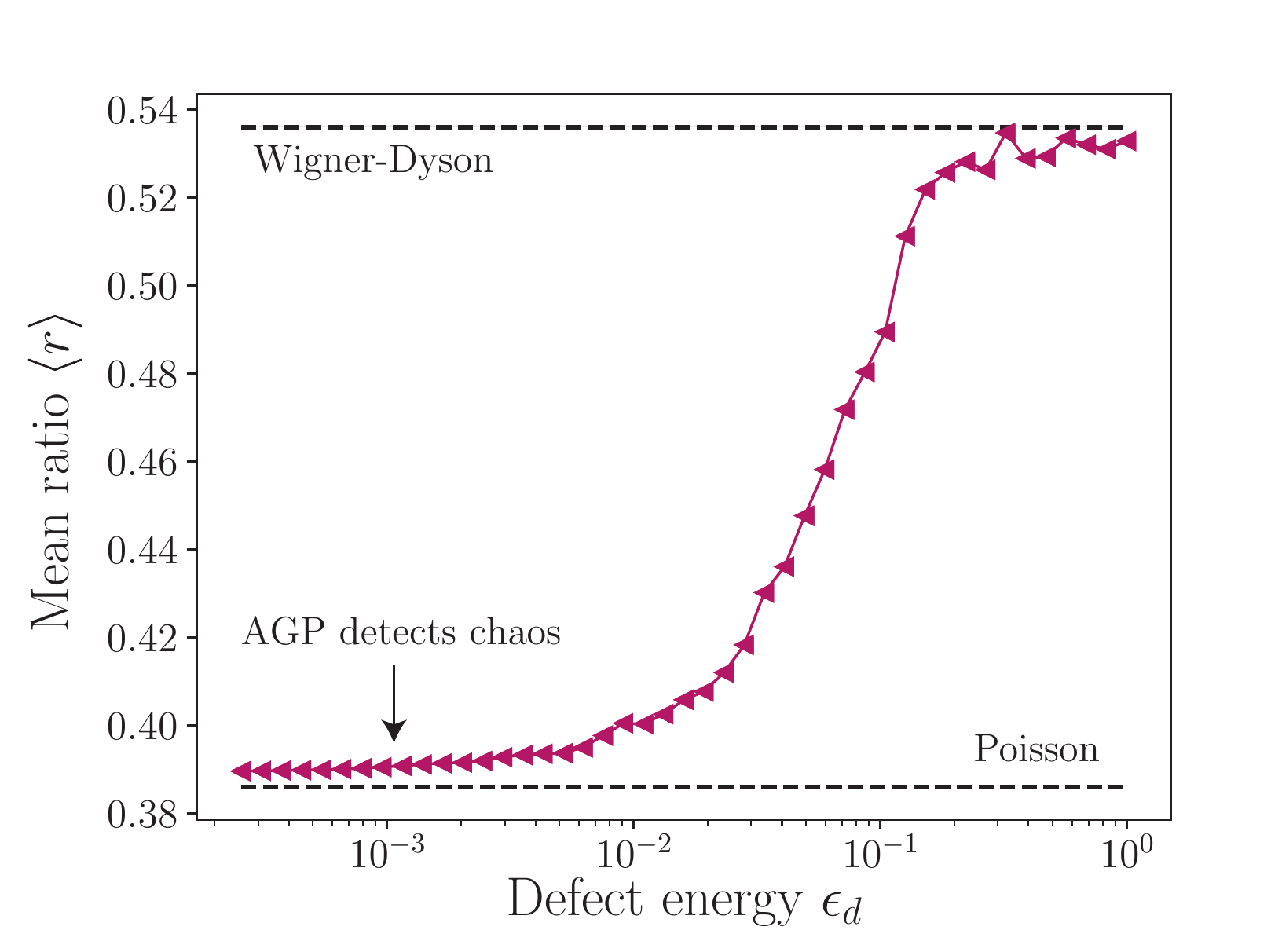}
\caption{\textbf{Energy level statistics.} Mean ratio of energy level spacings $\langle r \rangle$ as a function of defect energy $\epsilon_d$ for an XXZ chain of length $L=16$ at anisotropy $\Delta=1.1$. The arrow indicates the value of the defect energy where chaos can be detected (for $L=16$) using the exponential scaling of AGP norm.}
\label{XXZlevelstat}
\end{figure}

To contrast the scaling of the AGP norm with more traditional approaches in Fig. \ref{XXZlevelstat}, we show the mean ratio of energy level statistics as a function of defect energy for system size $L=16$. Given subsequent energy level spacings $s_n = E_{n+1}-E_{n}$, this ratio is defined as
\begin{equation}
r_n = \frac{\min(s_n,s_{n+1})}{\max(s_n,s_{n+1})}.
\end{equation}
For non-ergodic systems and Poissonian level statistics, $\langle r \rangle\approx 0.386$, whereas for chaotic systems and Wigner-Dyson statistics  $\langle r \rangle\approx 0.536$. In this model, the average ratio $\langle r \rangle$ shows the crossover from non-ergodic to ergodic behavior at $\epsilon_d^*\sim 0.1$ \cite{chavda2014poisson}. This crossover value of $\epsilon_d$ has a very weak dependence on the system size. In comparison, for the same system size $L=16$ the AGP norm shows a clear crossover to chaos for a much smaller $\epsilon_d^* \sim 10^{-3}$ (see Fig. \ref{defect_XXZ} a) ). For larger system sizes, the gap between the chaos thresholds extracted by these two methods becomes even larger. Moreover, we also estimated the critical perturbation strength using the spectral form factor for the same system size $L=16$. Since this generally doesn't self-average \cite{prange1997spectral,braun2015self}, we added disorder to the $zz$-coupling in the Hamiltonian (Eq.~\eqref{Ham_XXZ_defect}) which reduces the sensitivity of this probe to detect chaos. From the spectral form factor we find $\epsilon_d^*\sim 0.1$, a value where the level statistics is roughly half way between Poisson and Wigner-Dyson (see Fig.~\ref{XXZlevelstat}). Such a correspondence was also observed for disordered models in Ref.~\cite{vsuntajs2019quantum}.

We believe that the reason that the AGP norm is so much more sensitive is that it effectively detects the change in the differential of the norm with the system size. The absolute value of the AGP norm at the threshold is still much closer to the integrable value than to the chaotic one. Such a differential is much harder to detect using other measures, e.g. the level spacing ratio, because this crossover is much smoother, and it is harder to define a sharp threshold.

\begin{figure}[ht]
	\centering
\includegraphics[width= 0.45\textwidth]{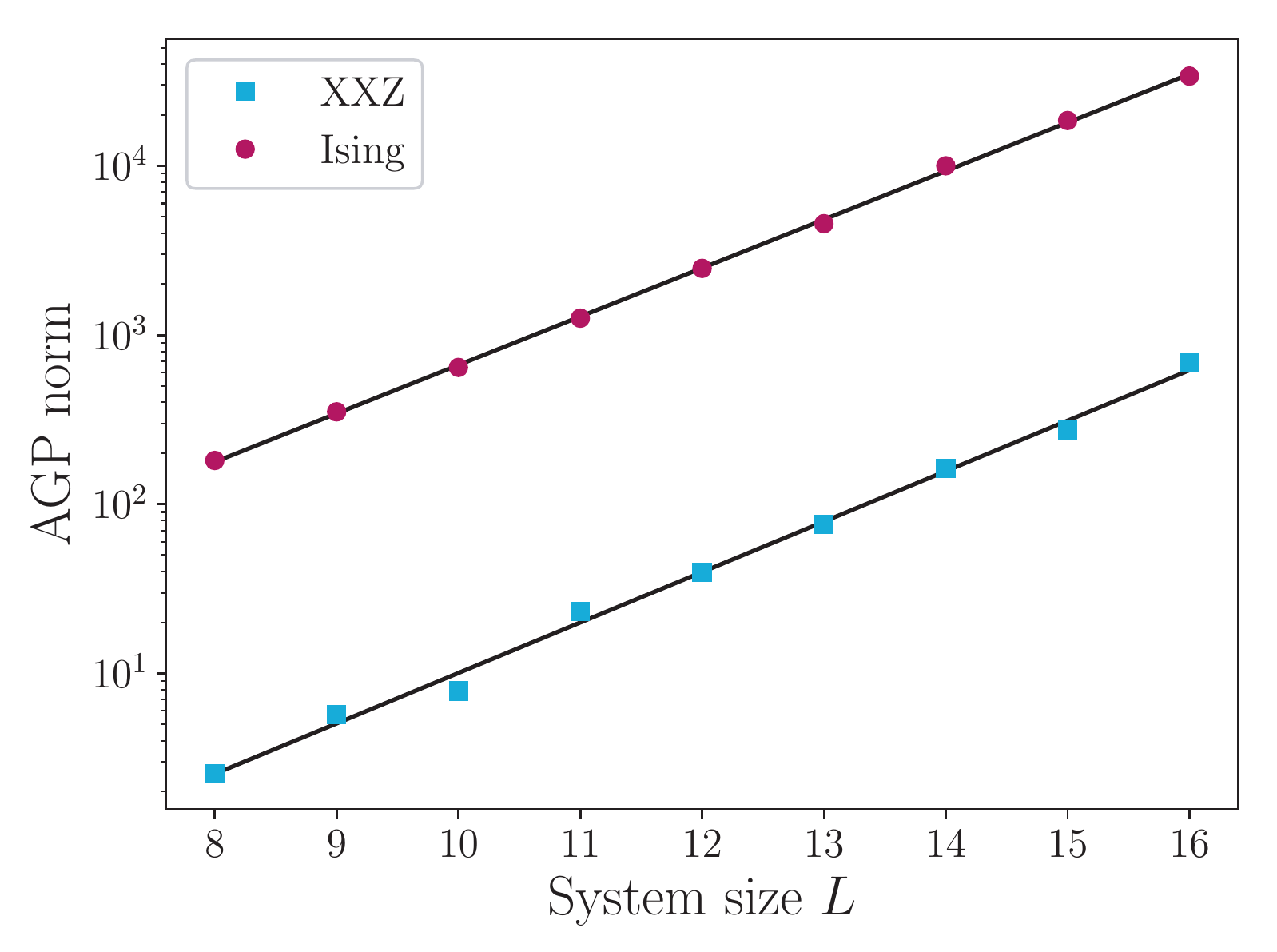}
\caption{\textbf{Integrability-breaking deformation at the integrable point.} The AGP norm $||\mathcal{A_\lambda}||^2$ shows an exponential scaling at the integrable point for the XXZ chain (squares) with $\lambda=\epsilon_d$ and the Ising chain (circles) with $\lambda=h_z$. The black lines correspond to  exponential fits, i.e. $||\mathcal{A_\lambda}||^2 \sim e^{ \beta L}$, where $\beta \approx \log (2)$. \textit{XXZ Parameters:} $\Delta=1.1, \epsilon_d=0$, \textit{Ising Parameters:} $h_x=0.75, h_z=0$.}
\label{defect_XXZ_B}
\end{figure}


In Fig.~\ref{defect_XXZ_B} we show similar results, now choosing to deform the Hamiltonian in the direction of the integrability-breaking operator itself, i.e. $\lambda=\epsilon_d$ for the XXZ chain and $\lambda=h_z$ for the Ising chain. We choose to work in the full Hilbert space with dimension $\mathcal{D}=2^L$. We find that the AGP norm shows exponential scaling even when $\epsilon_d=0$, i.e. when the Hamiltonian is integrable. We find a good fit to the exponential scaling $||\mathcal{A}_\lambda||^2 \sim e^{\beta L}$, with now $\beta \approx \log(2)$. Again we confirm that the results remain the same if we use an extensive integrability-breaking term instead~(see Appendix~\ref{append.universalslope}).


\section{Long relaxation times}
We already mentioned a very peculiar fact following from Fig.~\ref{defect_XXZ}: namely, instead of the perhaps expected crossover of the integrable polynomial scaling of the AGP norm to the ETH exponential scaling with the slope $\log(2)$ the AGP crosses over to the exponential scaling regime with the slope $\beta=1.28$, which is almost twice as large as the slope predicted by ETH, $\beta=\log (2)\approx 0.69$.  Combining this result with Eq.~\eqref{eq:norm_agp_f_omega}, which we highlight works in both integrable and nonintegrable regimes, we conclude that at small $\omega$ the function $|f_\lambda(\omega)|^2$ should scale exponentially with the system size. This implies that the system must have exponentially long relaxation times, which are known to exist in classical chaotic systems like the FPUT chain \cite{gallavotti2007fermi,danieli2017intermittent, pace2019behavior}. Although we cannot rule out the eventual relaxation to the ETH value for system sizes greater than those we have studied, our results here suggest that, while an exponentially small perturbation is sufficient to induce chaos in the system, it takes an exponentially long time for the system to relax to the steady state.  In Appendix \ref{append.NNN}, we show that a similar behavior persists if we break the integrability by a small extensive perturbation, here chosen as the second nearest-neighbor Ising interactions. We found the same slope of $\beta\approx 1.28$, ruling out that this scaling is induced by the ultra-local nature of the perturbation in Fig.~\ref{defect_XXZ}~a). As the defect energy is increased further to large values (in particular,  $\epsilon_d\sim 1$), we find that the slope of AGP norm's exponential growth reduces again to the ETH value of $\beta\approx \log (2)$ (see Appendix~\ref{append.universalslope}). 

To make the connection between the AGP norm and the relaxation time more explicit let us observe that from Eq.~\eqref{eq:norm_agp_f_omega} for sufficiently small $\mu$ one can make the following estimate:
\begin{equation}
||\mathcal{A_\lambda}||^2 \sim \frac{|f_\lambda(\mu)|^2}{\mu}.   
\end{equation}
For integrable directions $\lambda$ (e.g. $\lambda=\Delta$ for the XXZ model) and $L> L^\ast$, where the AGP norm has exponential scaling, the norm becomes 
\begin{equation}
||\mathcal{A_\lambda}||^2 \sim C e^{\beta (L-L^\ast)},
\end{equation}
where $C$ roughly is the value of the unperturbed AGP norm at $L^\ast$. Recall that we observed a scaling of the critical perturbation strength like $\epsilon_d \sim e^{-\alpha L^\ast}$, such that one finds 
\begin{align}
|f_\lambda(\mu)|^2 &\sim C\mu e^{\beta (L-L^\ast)}\sim C \epsilon_d^\eta e^{\kappa L}, 
\end{align}
where $\eta=\beta/\alpha$, and $\kappa=\beta-\log(2)$, and we have neglected all polynomial factors in system size. For the XXZ model, the exponents are $\eta\approx 1.6$ and $\kappa \approx 0.85 \, \log(2)$ (see caption of Fig.~\ref{defect_XXZ}). Because $|f_\lambda(\omega)|^2$ is the Fourier transform of the two-point correlation function of $\partial_\lambda H$ (see~\eqref{eq:f_omega_def}), as $\omega\to 0$ it is proportional to the relaxation time of the system. Combining these considerations, we see that for the XXZ model we have
\begin{equation}
\tau \sim \epsilon_d^\eta e^{\kappa L},
\end{equation}
with both $\kappa$ and $\eta$ of $O(1)$. Similarly, for the Ising model, $\tau \sim h_z^\eta e^{\kappa L}$ where $\eta\approx 1.8$ and $\kappa \approx 1.28 \, \log(2)$ (see caption of Fig.~\ref{defect_XXZ}). We see that the relaxation time increases exponentially with the system size. For large system sizes it can saturate at some $L$-independent value, which should diverge as $\epsilon_d\to 0$. This would reflect the crossover of the scaling of the AGP norm to the ETH result: $||A_\lambda||^2\propto \exp[S(L)]=\exp[\log(2) L]$. While this scenario seems likely, we do not see any signatures for such a crossover within our numerics and thus cannot rule out more exotic scenarios for the behavior of the relaxation time with the system size. Moreover, at intermediate system sizes accessible to our numerics, we see an extremely stable exponential scaling of the AGP norm (and hence of the relaxation time), with the exponent $\beta$ independent of the strength of the integrability-breaking perturbation as long as it is sufficiently small. 
Interestingly, in a follow up work~\cite{tamiro_2020} a similar exponential scaling of the AGP norm with $\beta\approx 2
\log(2)$ was observed in a disordered central spin model even in the absence of any small parameters, i.e. at large integrability-breaking perturbations. 
We note that in all the systems analyzed so far in this regime, $\beta$ saturates near the maximum allowed value $2\log(2)$, within numerical precision. From the point of view of operator spreading, this value is very reminiscent to the $2\log(2)$ scaling of the operator entanglement entropy in maximally chaotic dual-unitary models~\cite{Bertini_2020}. Whether it is a simple coincidence or there is a deeper connection remains to be understood.

To illustrate these general considerations about the relaxation times we extracted the function $|f_\lambda(\omega)|^2$ directly. Usually it is very difficult to do so at exponentially small frequencies of interest, since there are very few eigenstates involved, hence leading to large fluctuations.
Here we computed $|f_\lambda(\omega)|^2$ by replacing all the delta-functions in Eq.~\eqref{eq:f_omega_def} with Lorentzians of width $\mu$. In all the figures $\mu=L2^{-L}$, consistent with the AGP regularization. The total spectral weight was subsequently computed on a logarithmically spaced grid. All the figures show the average spectral weight in each bin.

\begin{figure}[ht]
	\centering
\includegraphics[width= 0.45\textwidth]{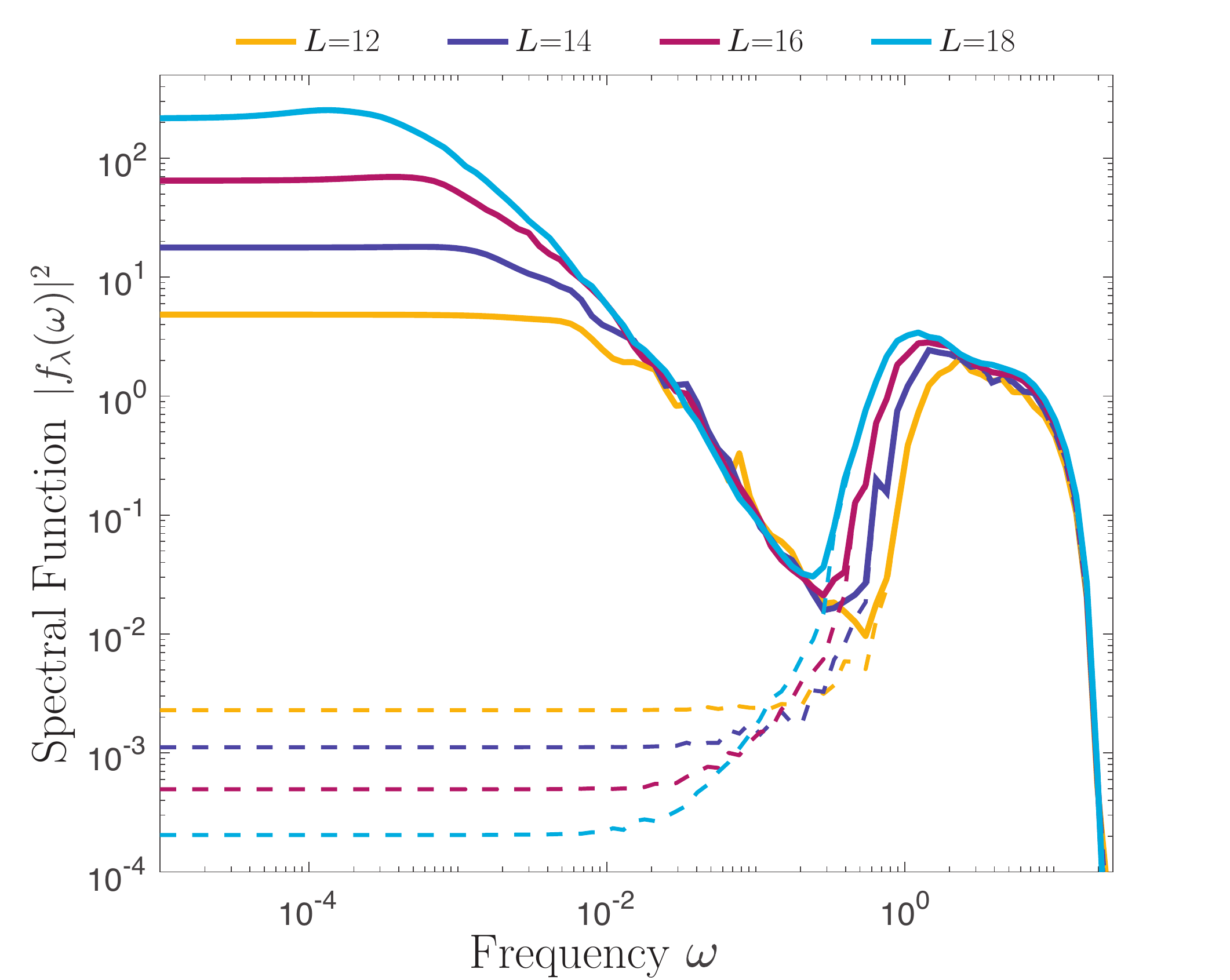}
\caption{\textbf{The spectral weight for the integrable perturbation.} The spectral weight $|f_\lambda(\omega)|^2$ for the integrable perturbation $\lambda=\Delta$ in the XXZ model at small integrability breaking perturbation $\epsilon_d=0.05$ (solid lines) and at the integrable point $\epsilon_d=0$ (dashed lines). The remaining parameters are the same as in Fig.~\ref{defect_XXZ}.}
\label{spectral_XXZ}
\end{figure}

In Fig.~\ref{spectral_XXZ} we show the extracted spectral weight $|f_\lambda(\omega)|^2$ for the XXZ model with $\lambda=\Delta$ for a small integrability breaking perturbation $\epsilon_d=0.05$ (solid lines) and exactly at the integrable point $\epsilon_d=0$ for four different system sizes $L=12,14,16,18$. As predicted from the AGP scaling, there is a clear exponentially growing spectral weight at small frequencies with an exponentially shrinking frequency range, where it plateaus. In the integrable regime, conversely $|f_\lambda(\omega)|^2$ is exponentially decreasing with the system size, approaching zero in the thermodynamic limit.

\begin{figure}[ht]
	\centering
\includegraphics[width= 0.45\textwidth]{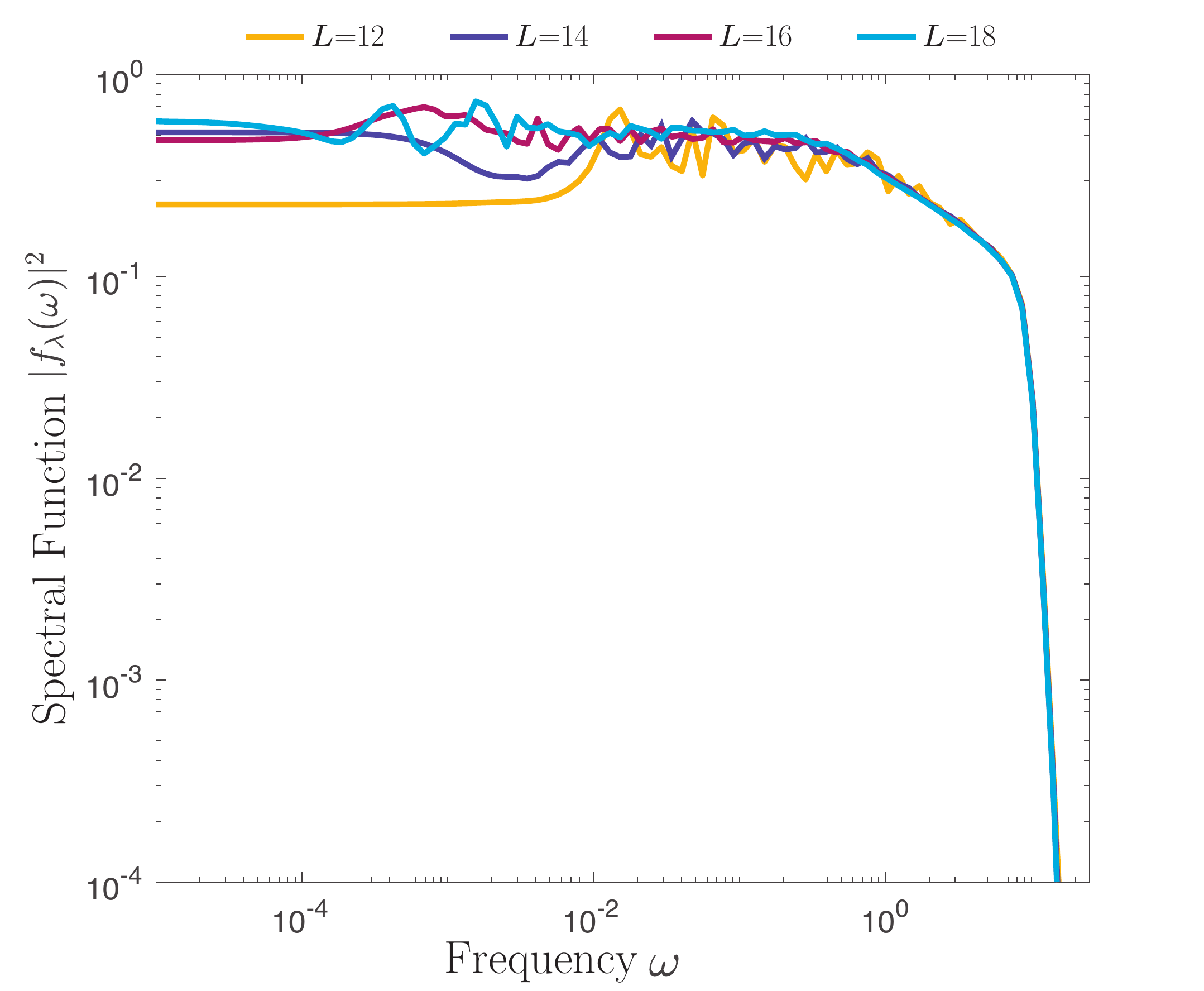} \\
\includegraphics[width= 0.45\textwidth]{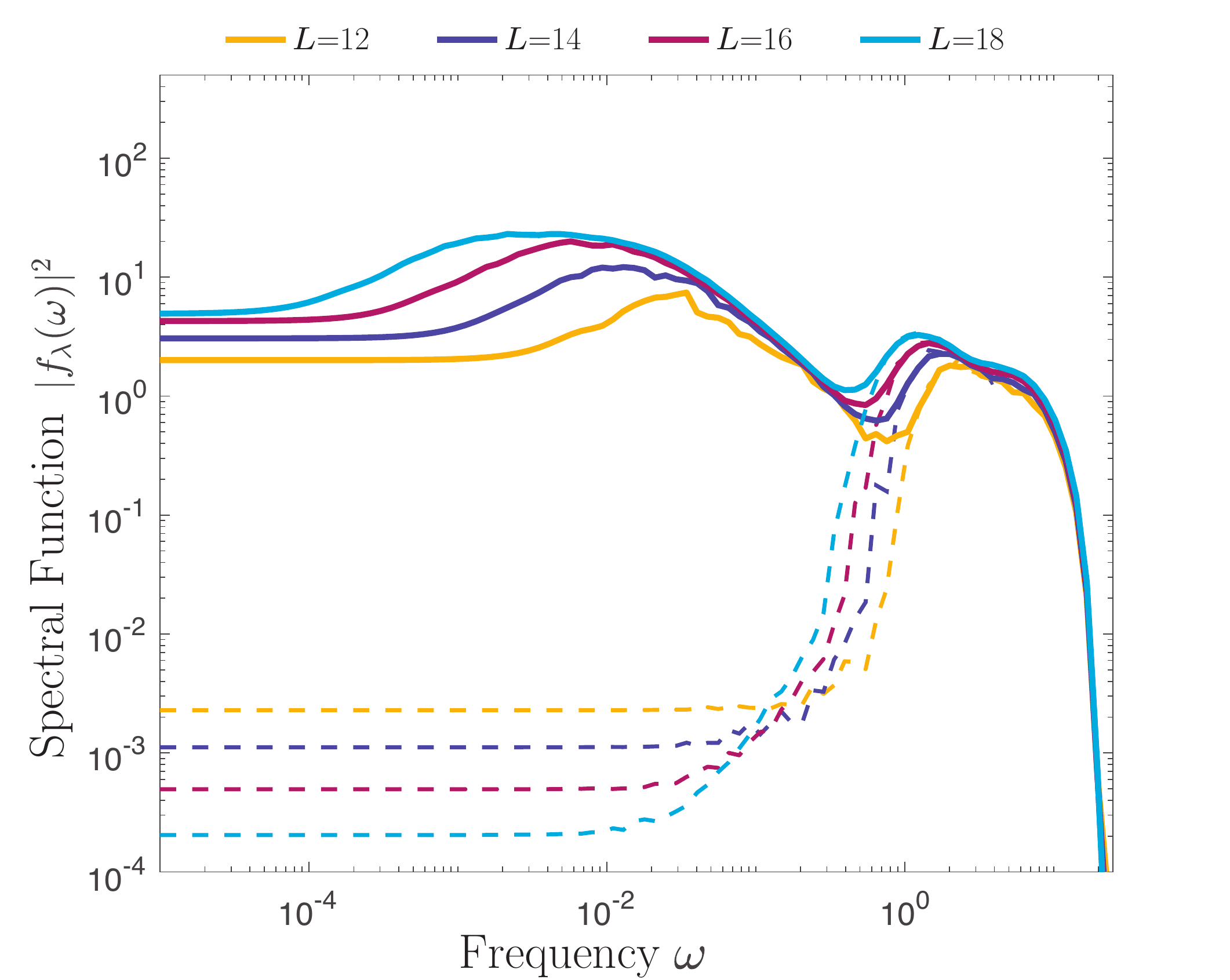}
\caption{\textbf{The spectral weights for the non-integrable perturbation.} The spectral weight $|f_\lambda(\omega)|^2$ 
for the non-integrable perturbation $\lambda=\epsilon_d$ in the XXZ model at the integrable point $\epsilon_d=0$ (top) and for the perturbation $\lambda=\Delta$ at the strongly non-integrable point, i.e. in the ETH regime, $\epsilon_d=0.5$ (bottom). The remaining parameters are the same as in Fig.~\ref{spectral_XXZ}.}  
\label{spectral_XXZ_1}
\end{figure}


To contrast this behavior of the spectral function with the other two regimes where the AGP norm shows exponential scaling with $\beta=\log(2)$, in Fig.~\ref{spectral_XXZ_1} we show $|f_\lambda(\omega)|^2$ in such regimes. The top plot shows the $|f_\lambda(\omega)|^2$ for the nonintegrable perturbation $\lambda=\epsilon_d$ at the integrable point of the XXZ model $\epsilon_d=0$. While the bottom plot corresponds to the perturbation $\lambda=\Delta$ at the strongly nonintegrable point $\epsilon_d=0.5$ where the system satisfies ETH~\cite{gubin2012quantum, brenes_2020}.



\section{Distinguishing between integrable and ETH regimes }

The AGP clearly depends on both the Hamiltonian $H$ and the direction along which it is deformed, i.e. $\partial_{\lambda}H$. In the previous sections, we argued that generic perturbations in chaotic systems lead to an AGP norm scaling exponentially with system size, whereas in integrable models integrability-preserving perturbations lead to an AGP norm scaling polynomially. This scaling is directly reflected in the relaxation times of $\partial_{\lambda}H$ through its probing of the zero-frequency limit of $|f_{\lambda}(\omega)|^2$. However, in specific cases, polynomial scaling of the gauge potential can also be observed in chaotic systems.

In particular, there is a special class of operators which can be represented as $K=i [H,B]$, where $B$ is a local operator or a sum of local operators. A current can, e.g., be represented in this way as $B=\sum_i i\, n_i$, where $n_i$ is the conserved charge; $n_i=\sigma_z^i$ for the XXZ model. For such operators $\mathcal A_\lambda=B$ by construction, and the AGP will have a polynomial norm irrespective of whether the system is integrable or chaotic. For such operators $|f_\lambda(\omega)|^2$ must also vanish at $\omega \to 0$, consistent with recent numerical results~\cite{brenes_2020}. On a related note, see \cite{dymarsky2019new}. Physically, this non-divergence of the AGP, even in the chaotic systems satisfying ETH, simply follows from the fact that deforming the Hamiltonian with the operator $K$ is a symmetry transformation, which does not change the spectrum of the Hamiltonian, but simply transforms the eigenstates with the unitary operator $U=\exp(-i\lambda B)$. When checking for quantum chaos, such deformations can be explicitly excluded when probing the scaling of the gauge potential.

While the existence of nontrivial deformations with polynomial scaling of the AGP norm is an indicator of integrability, generic integrability-breaking perturbations give rise to exponential scaling, in which case the specific dependence on $\mu$ offers further information. Note that this also implies the existence of a family of integrable models, excluding more exotic `isolated' integrable systems where every possible perturbation breaks integrability.

In the previous section, the scaling of the AGP norm was the same as one would expect from ETH, even though at $\epsilon_d=0$ the system is integrable and ETH is clearly violated. The non ETH-behavior can be seen,e.g., in large eigenstate-to-eigenstate fluctuations of the expectation value of $\sigma^z_{{\lceil}(L+1)/2 {\rceil}}$~\cite{brenes_2020a}. For this perturbation the scaling of the AGP with the system size simply tells us that $|f_\lambda(\omega)|^2$, which remains well-defined in such models, saturates to a nonzero constant at small $\omega$, as confirmed directly in the previous section. Similar to the usual matrix elements of observables, the information about the integrability of the system is now contained in the statistical properties of the AGP norm.

More specifically, for random matrix ensembles the statistical properties of the fidelity susceptibility (equivalent to the contributions to the AGP norm for individual eigenstates) were analyzed in Ref.~\cite{Sierant_2019}, where the distribution for different eigenstates is considered. The fidelity susceptibility $z_{n,\lambda}$ of an eigenstate $|n(\lambda)\rangle$ is equivalent to
\begin{align}
z_{n,\lambda} \equiv \frac{1}{\mathcal{D}} {\langle n|\mathcal {A_\lambda}^2|n\rangle_c}  \equiv \frac{1}{\mathcal{D}} \sum_{m\neq n}|{\langle n|\mathcal {A_\lambda}|m\rangle|^2},
\end{align}
such that $||\mathcal{A}_{\lambda}||^2 = \sum_n z_{n,\lambda}$. 

Let us briefly present a simple derivation of the tail of this distribution and then contrast the AGP distribution for integrable and ETH regimes. The tail of this distribution for typical (random) perturbations will be dominated by contributions from neighbouring energy levels, such that its distribution will be inheriting its properties from the level spacing distribution.

Recall that the exact AGP norm with $\mu=0$ is given by Eqs.~\eqref{AGP_orign_def} and~\eqref{eq:L2_norm}. For a typical perturbation we can replace the numerator of Eq.~\eqref{AGP_orign_def} with a random matrix such that typical matrix elements are of order $1/\sqrt{\mathcal{D}}$ (see \eqref{eq:fwETH}). The tail of the distribution for large $z_{n,\lambda}$ is dominated by nearby energy levels and we can approximate
\begin{equation}
\label{eq:z_l}
z_{n,\lambda} \approx \frac{C}{s_n^2},
\end{equation}
where $s_n$ is the level spacing $E_{n+1}-E_{n}$ now normalized by the Hilbert space dimension (such that the mean value of $s$ is unity) and $C$ is an unimportant constant, which we can set to one. The scaling of the probability distribution at large $z_\lambda$ follows as
\begin{equation}
{\rm Pr} (z_\lambda=x) \sim  \frac{1}{x^{3/2}}P\left({1\over \sqrt{ x}}\right),
\end{equation}
where $P(s)$ is the normalized nearest-neighbour level spacing distribution. 

\begin{figure}[ht]
	\centering
\includegraphics[width= 0.45\textwidth]{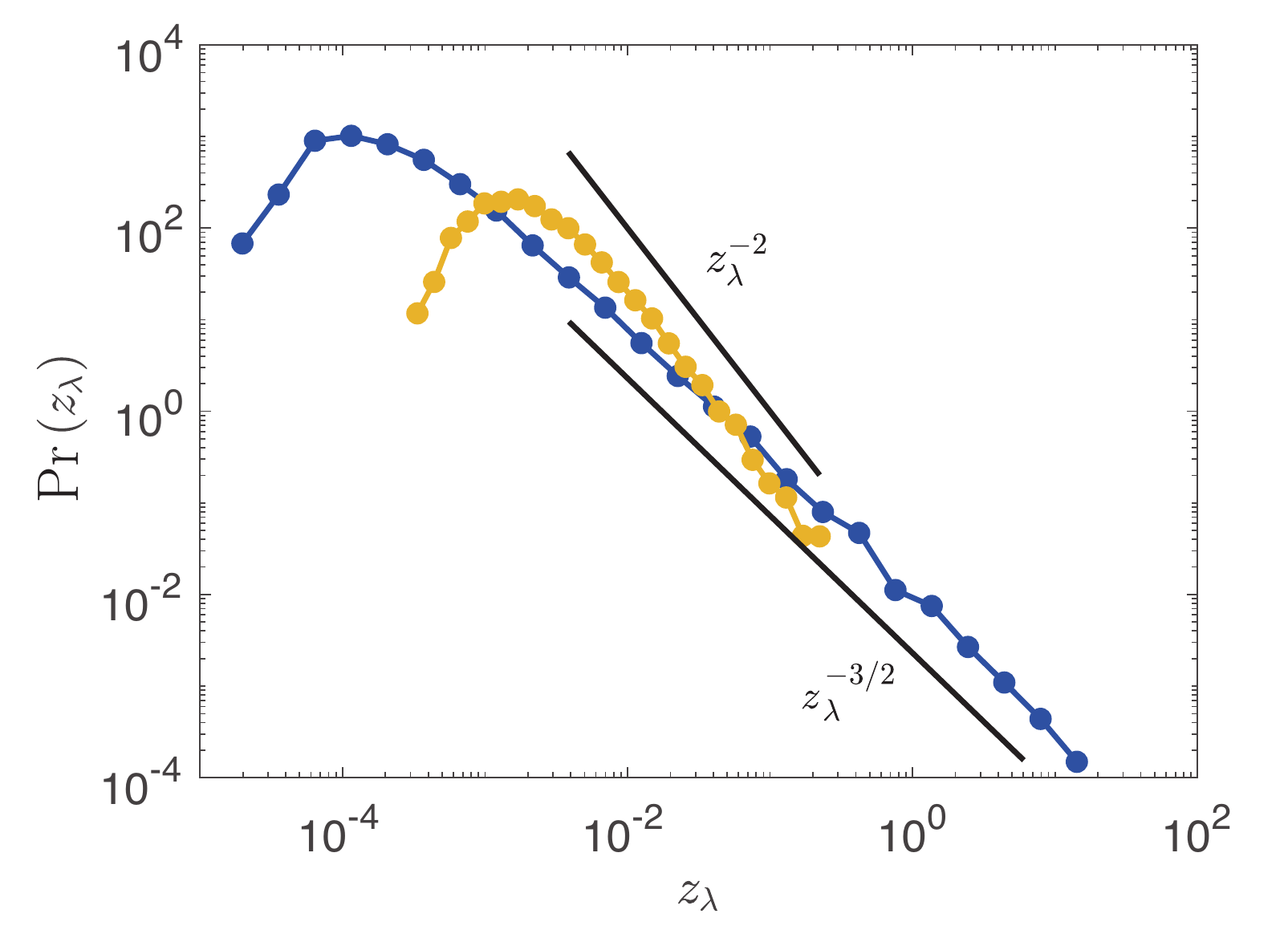}
\caption{\textbf{AGP norm distribution}. Distribution of the eigenstate contributions $z_\lambda$ to the rescaled AGP norm (see Eq.~\eqref{eq:z_l}) for the XXZ model with $L=16$ spins. The two curves describe the results for the non-integrable perturbation $\lambda=\epsilon_d$ at the integrable point $\epsilon_d=0$ (blue) and for the perturbation $\lambda=\Delta$ at the strongly non-integrable point $\epsilon_d=0.5$ (yellow). Black lines show the expected scalings $z_\lambda^{-3/2}$ and $z_\lambda^{-2}$ for the integrable and non-integrable model respectively.}
\label{distribution}
\end{figure}

For integrable systems there is no level repulsion, $P(s \to 0) \neq 0$, and we have (to dominant order)
\begin{equation}
 { \rm Pr} (z_\lambda=x) \propto {x^{-3/2}},
\end{equation} 
for $x\gg 1$. Note that, as a consequence of this fat tail, the mean AGP diverges without regularization. The regularization with $\mu$ in the norm of the AGP effectively introduces a cutoff to the energy denominator at the rescaled cutoff $\bar \mu= \mu \mathcal D$. Assuming that the AGP norm is dominated by the contributions $z_{n,\lambda}$ for which the derived scaling holds, we can say that the average fidelity susceptibility is given by $\left< z_\lambda \right> \propto 1/\bar \mu$, and hence $||\mathcal A_\lambda||^2=\mathcal D \left< z_\lambda \right>~\sim \mathcal D/\bar \mu$. This agrees with the observed scaling shown in  Fig.~\ref{defect_XXZ_B}. 
On the other hand, chaotic systems satisfying ETH exhibit level repulsion and $P(s) \approx s^\beta$, resulting in ${\rm Pr} (z_\lambda=x) \propto x^{-(3+\beta)/2}$ at large values of $x$. For the considered Ising and XXZ model, the relevant random matrix ensemble is Gaussian orthogonal ensemble (GOE), for which $\beta=1$ and
\begin{equation}
{\rm Pr} (z_\lambda=x) \propto x^{-2}.
\end{equation}
In contrast to the integrable model, the mean $\left< z_\lambda \right> \sim - \log( {\bar \mu}^2)$ diverges only logarithmically with the cutoff. These simple scaling arguments agree very well with numerical observations shown in Fig.~\ref{distribution}.


From this analysis, we can conclude that choosing a fixed $\mu\sim 1/\mathcal D$ leads to the same scaling of the AGP norm with the Hilbert space dimension for the integrable model with a chaotic deformation $\lambda$ and for the ergodic ETH model. However, these two limits can still be distinguished by either the different scaling of the AGP norm with the cutoff $\mu$ or, equivalently, by the presence of an exponential-in-system-size difference between the typical and the average contributions of individual states to the AGP norm in the former (integrable) regime and the lack of such exponential difference in the latter (ETH) regime.

\section{Conclusions}
We found that the properly-regularized norm of the adiabatic gauge potential, the generator of adiabatic deformations, can serve as an extremely sensitive probe of quantum chaotic behavior. Within chaotic systems, this norm scales exponentially with system size, whereas it scales polynomially in interacting integrable systems and is approximately system-size independent in non-interacting systems for adiabatic deformations preserving integrability. For adiabatic deformations breaking integrability, exponential scaling is generally observed.

Using the present method to investigate the effects of an integrability-breaking perturbation on the XXZ and Ising chains, we found that perturbations that are exponentially small in system size suffice to induce chaotic behavior. We also found that such a small integrability-breaking term leads to anomalously slow dynamics along the integrable directions, with the relaxation time scaling exponentially with system size. Such integrability-breaking perturbations can also be detected at the integrable point, where no anomalous dynamics occur. Even though typical perturbations show exponential scaling of the regularized norm of the adiabatic gauge potential, regardless of whether the system is integrable or not, one can distinguish the two cases by their dependence on the regularization parameter or by their fluctuations.

This motivates the use of the adiabatic gauge potential, which is connected with both deformations of eigenstates and operator dynamics, as a sensitive probe into either chaotic or integrable behavior of quantum many-body systems.

\section*{Acknowledgements}
Some of the numerical computations were performed using QuSpin \cite{weinberg2017quspin, weinberg2019quspin}. We would like to thank Marcos Rigol and Lea Santos for detailed and very useful feedback on the manuscript.
We also thank Anushya Chandran, Anatoly Dymarsky Lev Vidmar, Phil Crowley, Pranay Patil, Tamiro Villazon, Phil Weinberg and Jonathan Wurtz for useful discussions. We would also like to acknowledge technical support by Boston University's Research Computing Services. M.P and D.K.C acknowledge support from Banco Santander Boston University-National University of Singapore grant. P.W.C gratefully acknowledges support from a Francqui Foundation Fellowship from the Belgian American Educational Foundation (BAEF), Boston University's Condensed Matter Theory Visitors program, and EPSRC Grant No. EP/P034616/1. A.P. was supported by the NSF Grant DMR-1813499 and the AFOSR Grant FA9550-16- 1-0334. D.S acknowledges support from the FWO as post-doctoral fellow of the Research Foundation -- Flanders.

\begin{appendix}

\begin{figure*}[t]
	\centering
\includegraphics[width= 0.98\textwidth]{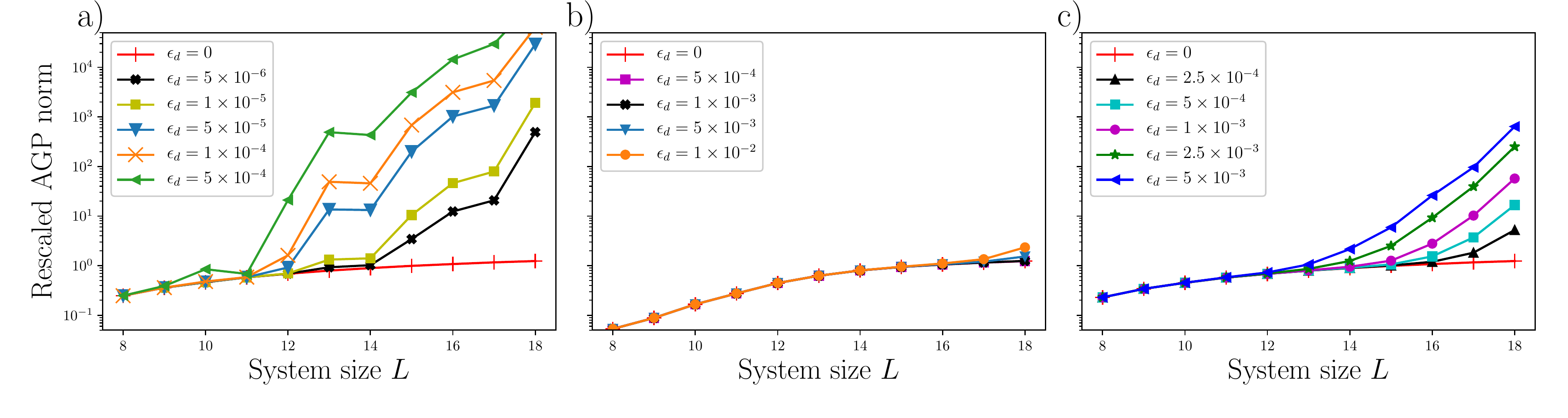}	
\caption{\textbf{Effects of regularization.} Size dependence of the rescaled AGP norm $||A_\lambda||^2/L$ for different choices of scaling for the cutoff $\mu$ close to chaotic-integrable transition point: a) When $\mu=L^{-1/2} \mathcal{D}_0^{-1}$, where $\mathcal{D}_0$ is the dimension of zero magnetization sector, the variation of the norm with system size is noisy.  b) When $\mu=L^2 \mathcal{D}_0^{-1}$, the norm, albeit smooth, is no longer very sensitive to small integrability-breaking perturbations and c) When $\mu=L \mathcal{D}_0^{-1}$, the norm is both appropriately smooth and exponentially sensitive to integrability-breaking perturbation   \textit{Model:} XXZ chain with defect in the middle (Eq.~\eqref{Ham_XXZ_defect}).  \textit{Parameters:}  $\Delta=1.1, \lambda=\Delta$  }
\label{append_fig_XXZ_small_mu}
\end{figure*}

\section{Cutoff scaling with system size} \label{append.muScaling}
Unless stated otherwise, in all calculations we have chosen a cutoff $\mu=L \mathcal{D}^{-1}$, where $\mathcal{D}$ is the dimension of the Hilbert space. The prefactor $L$ has been chosen to remove the logarithmic correction coming from the zero-frequency contribution of $|f(\omega=0)|^2 = L$ in chaotic models (see Appendix \ref{append.chaotic}). This can also be motivated by plotting the  AGP norm  and comparing it w.r.t. different choices of cutoff. We first study this norm close to chaotic-integrable transition point and then later describe its effect deep in the chaotic regime.

When we are close to the chaotic-integrable transition point and the cutoff is too small (e.g. $\mu=L^{-1/2}\mathcal{D}^{-1}$), then we find that the AGP norm is too sensitive to the exponentially close eigenstates, showing a non-smooth exponential scaling, which makes it hard to draw any conclusions (see Fig.~\ref{append_fig_XXZ_small_mu}~a) ). On the other hand, if the cutoff is too large (e.g. $\mu=L^2 \mathcal{D}^{-1}$), then the AGP norm, albeit smooth, is no longer sensitive to the small strength of integrability-breaking perturbation(see Fig.~\ref{append_fig_XXZ_small_mu}~b)). In Fig.~ \ref{append_fig_XXZ_small_mu}~c) with  $\mu=L \mathcal{D}^{-1}$ , we find that the rescaled AGP norm shows an exponential scaling that is both appropriately smooth and exponentially sensitive to integrability-breaking perturbations.  

Deep in the chaotic (ergodic) phase, we find that the numerically-obtained scaling for the norm of the AGP is almost the same for the different choices of cutoff scaling we studied. 
\section{Derivation of AGP for the free model}\label{append.free}
As shown in Refs.~\cite{del2012assisted, kolodrubetz2017geometry}, the AGP for changing the transverse field $h_x$ in a free Ising model with periodic boundary conditions is given by
\begin{equation}
\mathcal{A}_{h}= \sum_{l=1}^{L} \alpha_l O_l, 
\end{equation}
where the operators $O_l$ are given by the following Pauli string operators
\begin{equation}
O_l=  \sum_{j=1}^L ( \sigma_j^x \sigma_{j+1}^z \ldots \sigma_{j+l-1}^z \sigma_{j+l}^y +  \sigma_j^y \sigma_{j+1}^z \ldots \sigma_{j+l-1}^z \sigma_{j+l}^x),
\end{equation}
and the coefficients $\alpha_l$ are given by
\begin{equation}
\alpha_l= -\dfrac{1}{4 L} \sum_{k=0}^{\pi(L-1) /L} \dfrac{\sin(k) \sin(lk)}{(\cos k - h_x)^2 + \sin^2 k}.
\end{equation}
The norm of the AGP follows as
\begin{align}
||\mathcal{A}_{h}||^2 = \dfrac{1}{2^L} \Tr \left[ \mathcal{A}_{h}^2\right] =2 L   \sum_{l=1}^{L}  \alpha_l ^2, 
\end{align}
where $\Tr \left[O_l  O_p\right] =2^{L+1}L $ was used since all strings of Pauli matrices are trace-orthogonal. The above expression was used to compute the AGP norm for the free model in Fig.~\ref{exact_regim_norm} in the main text.

To obtain the scaling with system size, we can use the analytical expressions of $\alpha_l$ for large enough system sizes \cite{kolodrubetz2017geometry}, i.e. $\alpha_l=h_x^{-l-1}$ in the paramagnetic phase where $h_x^2 >1$.  Using this, we find that
\begin{align}
||\mathcal{A}_{h}||^2 &\sim \frac{1}{h_x^2 (h_x^2-1)} L (1-e^{-2L \log h_x }).
\end{align}
Recall that the correlation length in the transverse field Ising model $\sim 1/\log h_x$.

\section{AGP bound} \label{append.chaotic}
Recall that the norm of the AGP can be expressed as
\begin{eqnarray}
||\mathcal{A}_\lambda||^2 &=&  \int d \omega \, \dfrac{\omega^2}{ (\omega^2 + \mu^2)^2}  \overline{|f_\lambda(\omega)|^2}, 
 \label{eq:appC_normA}
\end{eqnarray}
with 
\begin{equation}
\overline{|f_\lambda(\omega)|^2}={1\over \mathcal D} \sum_n \sum_{m\neq n} |\langle n | \partial_\lambda H| m\rangle|^2 \delta(\omega_{nm}-\omega),
\end{equation}
and $\omega_{nm}=E_n-E_m$. It follows directly from eq.~\eqref{eq:appC_normA}, and $x^2/(x^2+1)^2\leq 1/4$,
that
 \begin{equation}
 ||\mathcal{A}_\lambda||^2\leq \frac{1}{4\mu^2} \int d \omega \,  |f_\lambda(\omega)|^2 = \frac{||\partial_\lambda H||^2}{4\mu^2}
 \label{eq:appCbound}
 \end{equation}
 Consequently, for any local perturbation the norm of the regularized AGP -- where we set $\mu \sim L 2^{-L}$ -- can't grow faster than $4^L$. Not only does it appear that this bound is saturated when probing integrable direction $\partial_\lambda H$ in models in which the integrability is weakly broken, it further implies that those observables $\partial_\lambda H$ take exponentially long to relax. Indeed, the above scaling can only be achieved by effectively having $|f_\lambda(\mu)|^2 \sim 2^L$. Yet, the total spectral weight, $\int d \omega \,  |f_\lambda(\omega)|^2$, is only polynomially large in the system size, implying that the corresponding spectral weight must be localized in a region $\Delta \omega \sim 2^{-L}$. Combined with expression~\eqref{eq:f_omega_def}, the latter implies $\partial_\lambda H(t)$ takes exponentially long to relax to equilibrium. 

For interacting integrable models we found $||\mathcal{A}||^2 \sim L^{\beta}$, where the exponent $\beta$ is non-universal. Since the norm is not exponential in system size, the function $|f_\lambda(\mu)|^2 \sim 2^{-L}$. This means that the function should vanish in the zero frequency limit, which implies  oscillatory dynamics of the observable $\partial_\lambda H(t)$.


\section{Effects of the anisotropy in the XXZ model.} \label{append.XXZ}
\begin{figure}[h]
	\centering
\includegraphics[width= 0.48\textwidth]{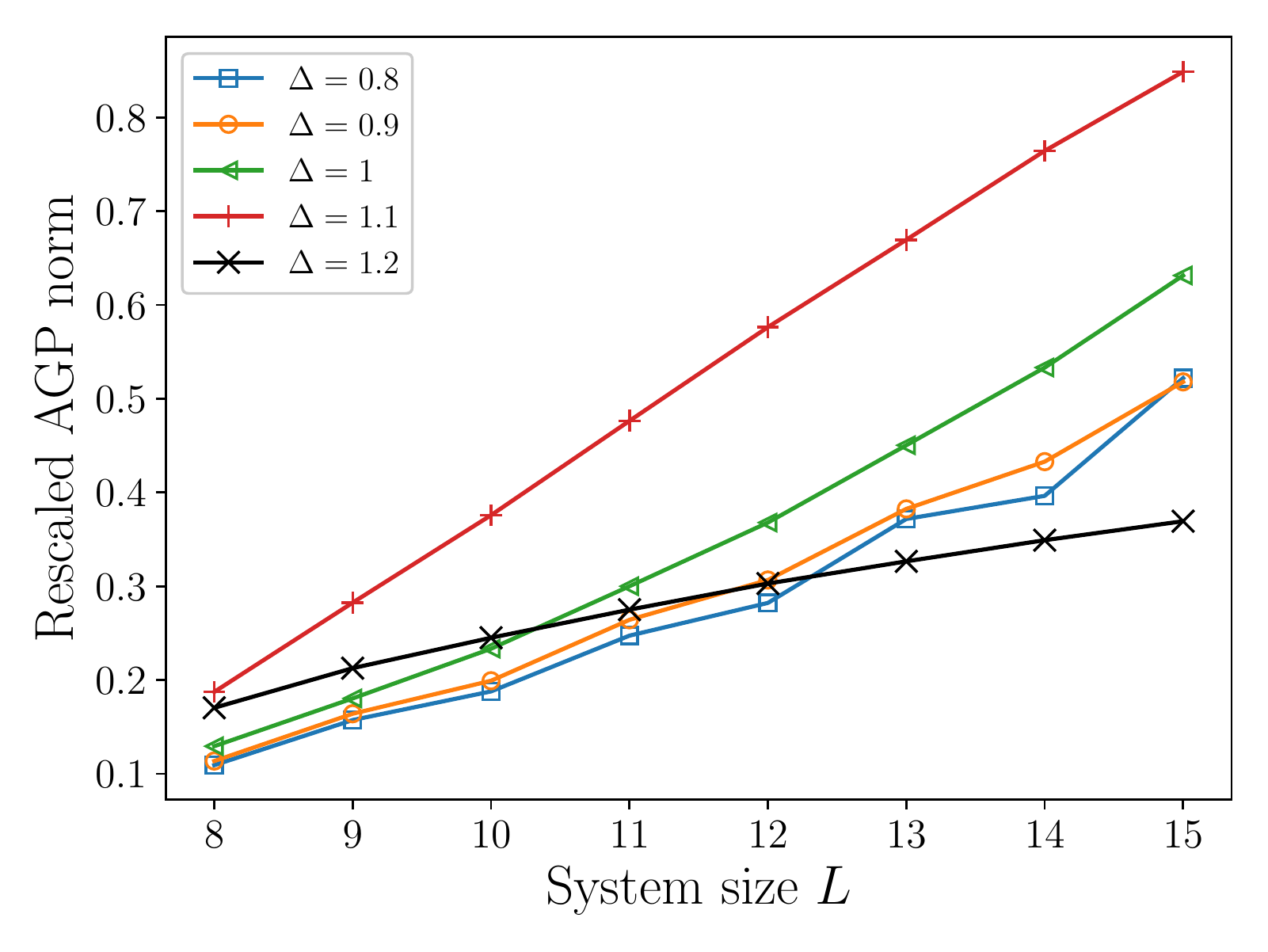}
\caption{\textbf{Anisotropy.} Rescaled AGP norm $||\mathcal{A}_\Delta||^2/L$ for the XXZ chain at different values of the anisotropy $\Delta$. }
\label{append_XXZ}
\end{figure}

In this Appendix, we will again consider the XXZ Hamiltonian (Eq.~\eqref{Ham.XXZ}):

\begin{equation}
H_{\text{XXZ}}=\sum_{i=1}^{L-1} ( \sigma_{i+1}^x \sigma_{i}^x + \sigma_{i+1}^y \sigma_{i}^y) + \Delta \sum_{i=1}^{L-1} \sigma_{i+1}^z \sigma_{i}^z, 
\label{Ham_XXZ_append}
\end{equation}
where $\Delta$ is the anisotropy, and we take $\Delta=\lambda$ as the adiabatic deformation, but now at different values of $\Delta$. We find that the slope of the AGP norm depends non-trivially on $\Delta$ (Fig. \ref{append_XXZ}). 

\section{NNN interactions in the XXZ chain} \label{append.NNN}

\begin{figure}[ht]
	\centering
\includegraphics[width=0.47\textwidth]{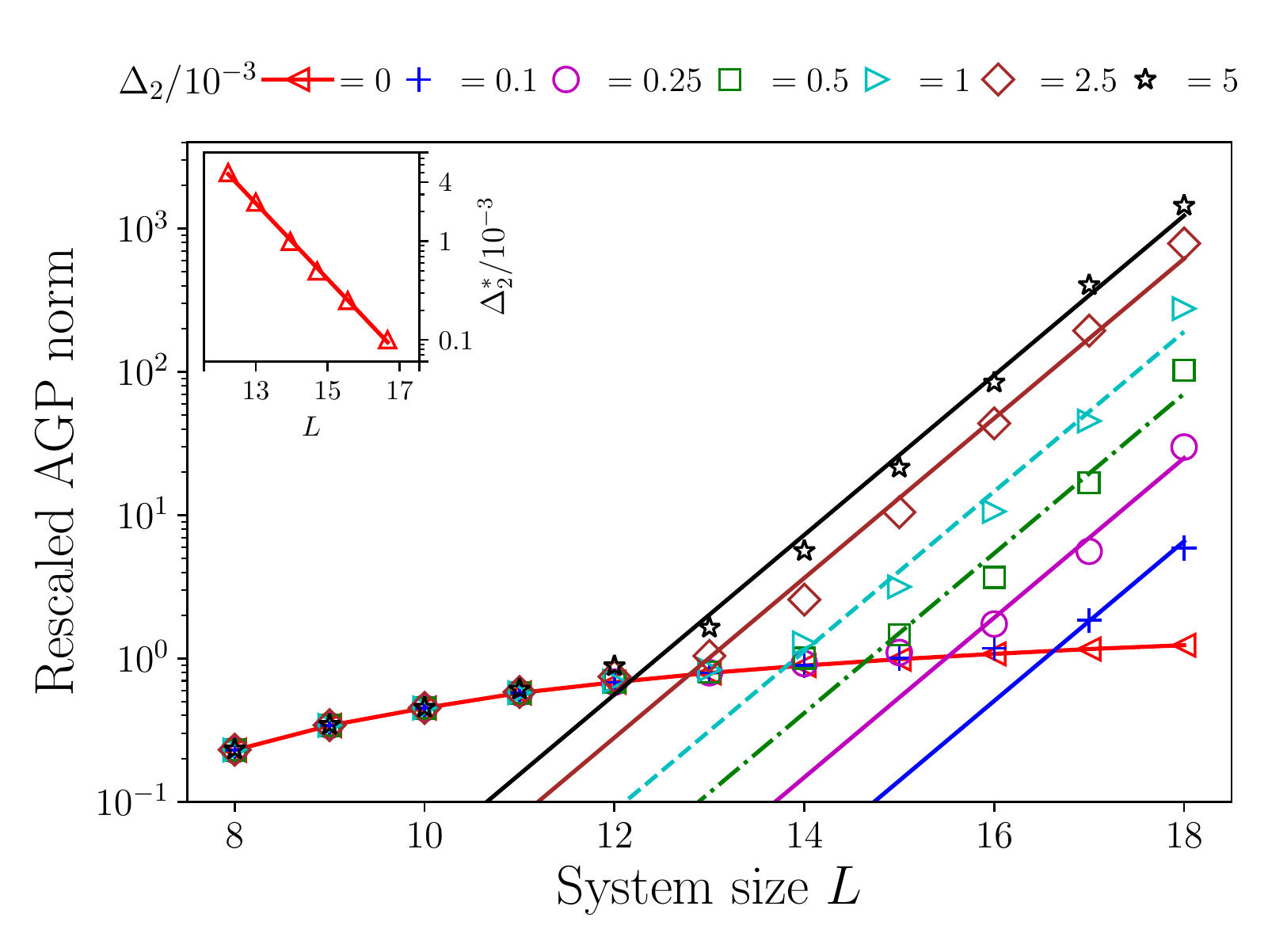}
\caption{  \textbf{Integrability breaking through NNN interaction:} Rescaled AGP norm $||\mathcal{A_\lambda}||^2/L$ with $\lambda=\Delta$ of the XXZ chain at $\Delta=1.1$ shows a sharp crossover from polynomial to exponential scaling with system size, even for very small perturbation strengths $\Delta_2$. As $\Delta_2$ decreases, the system size where this crossover happens increases. Straight lines are the exponential fits with $|A_{\lambda}||^2/L \sim e^{1.28 L}$.  \textit{Inset:} The integrability-breaking defect energy scales exponentially with system size, i.e. $\Delta_2^* \sim e^{-0.9 L}$. This is calculated for the symmetry sector with zero magnetization.}
\label{global_pertubrb_NNN1}
\end{figure}

\begin{figure}[ht]
	\centering
\includegraphics[width= 0.47\textwidth]{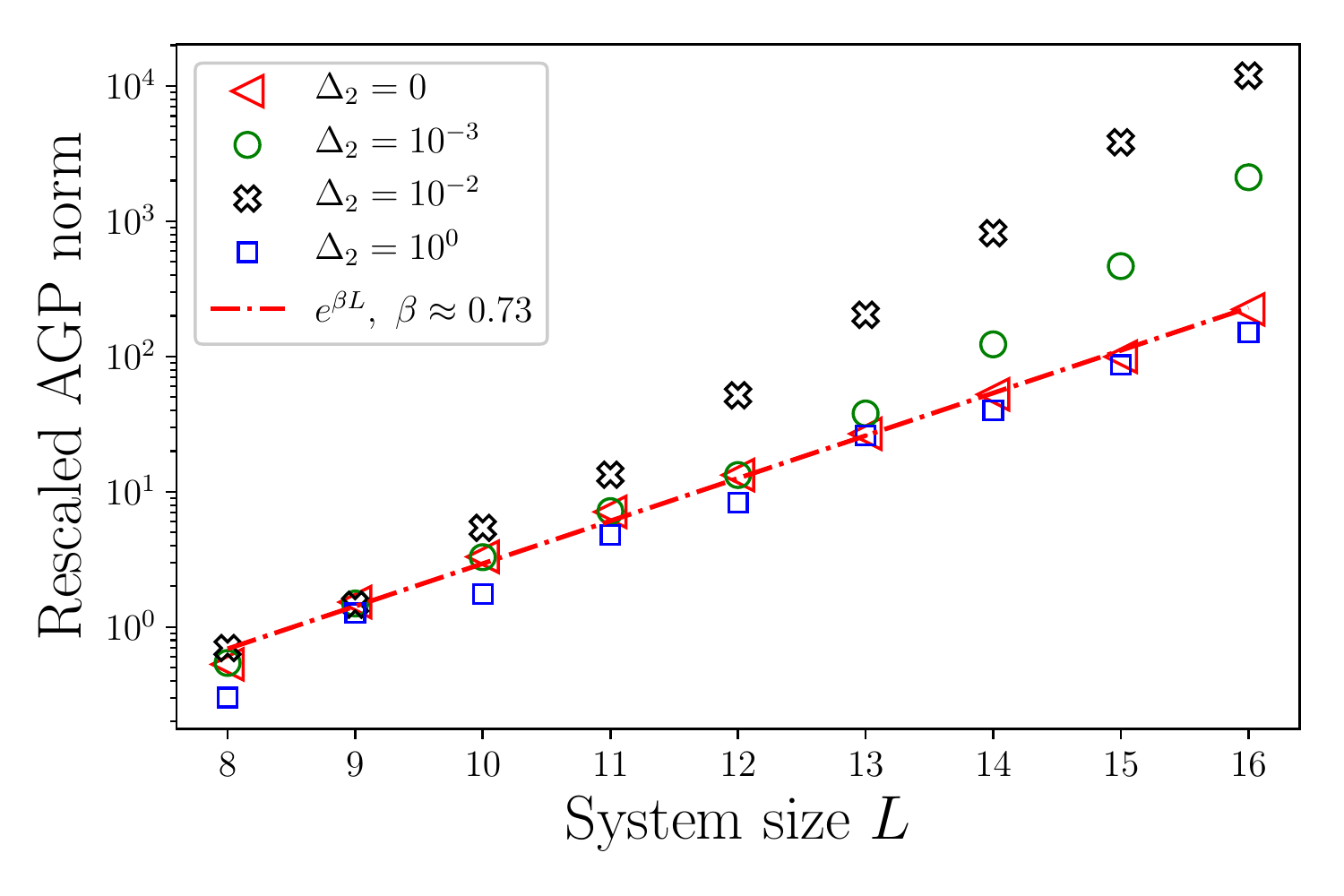}
\caption{\textbf{Integrability-breaking deformation.} Rescaled AGP norm $||\mathcal{A}_{\lambda}||^2/L$ for the XXZ chain at $\Delta=1.1$ with $\lambda=\Delta_2$. This is calculated for the full Hilbert space, not in any specific symmetry sector.}
\label{global_pertubrb_NNN2}
\end{figure}

In the main text, we studied the effect of strictly local integrability-breaking operator (whose support is a single site). Looking into the effects of the locality, we here study an extensive integrability-breaking operator. We add a next-nearest-neighbor (NNN) interaction to the XXZ chain, with Hamiltonian given as:
\begin{equation}
H_{\text{NNN}}=H_{\text{XXZ}}+ \Delta_2\sum_{i=1}^{L-2} \sigma^z_{i+2}  \sigma^z_{i}  , 
\label{NNN_Ham}
\end{equation}
The above model is chaotic for large enough $\Delta_2$ \cite{gubin2012quantum}. We choose $\lambda=\Delta$ (Fig. ~\ref{global_pertubrb_NNN1}) and  $\lambda=\Delta_2$ (Fig. ~\ref{global_pertubrb_NNN2}). In the limit $\Delta_2 \rightarrow 0$, when the above Hamiltonian (eqn. \ref{NNN_Ham}) is integrable,  the former (latter) is the integrability-preserving (breaking) direction.  As shown in Figs.~\ref{global_pertubrb_NNN1} and \ref{global_pertubrb_NNN2}, results are similar as for the strictly local perturbation studied in the main text. This implies our results are robust to the nature of the adiabatic deformation. 

\section{Universal slope of the AGP norm} \label{append.universalslope}
\begin{figure}[!htb]
	\centering
\includegraphics[width= 0.47\textwidth]{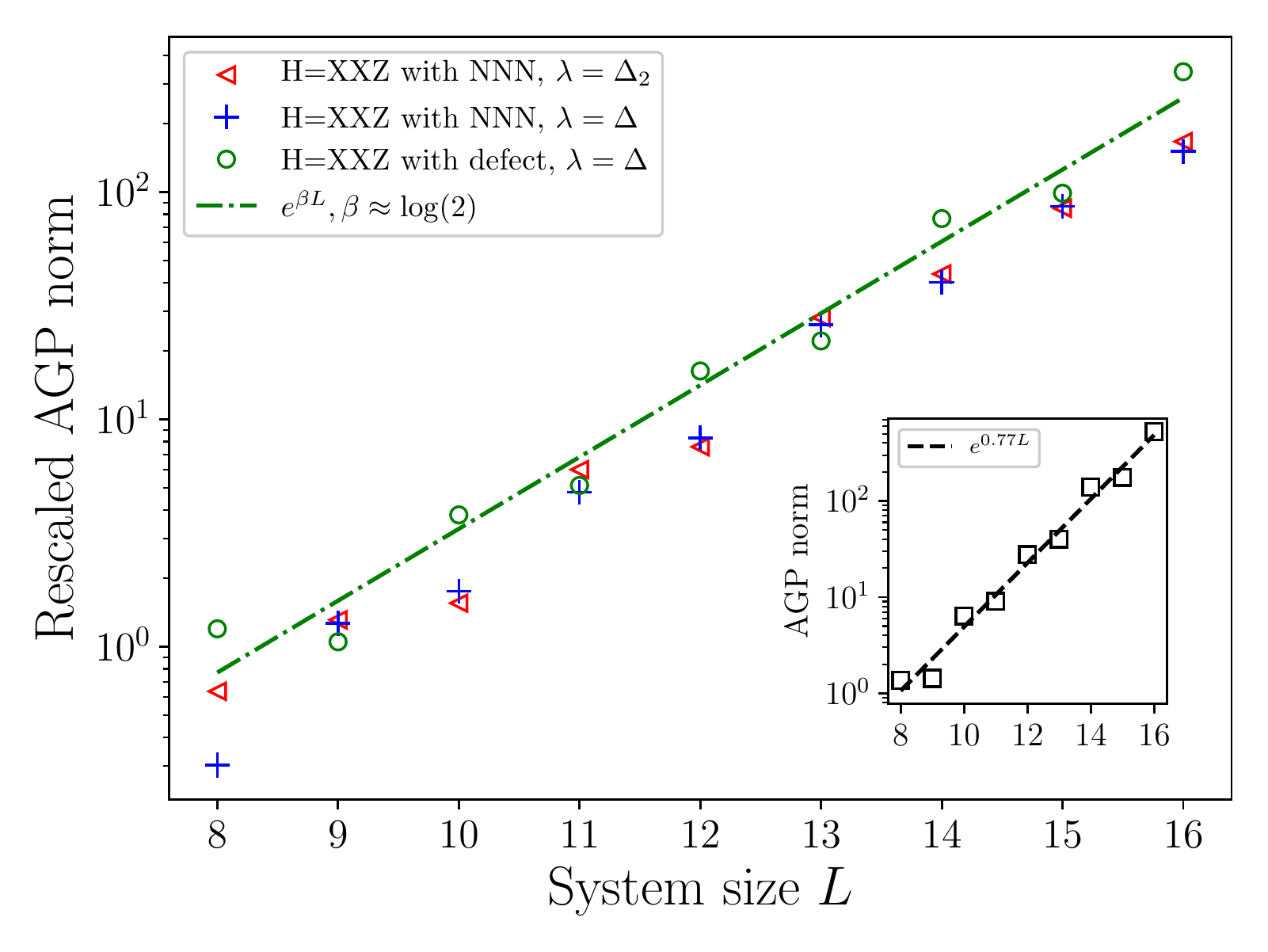}
\caption{\textbf{Universal slope at large integrability-breaking strengths.}  Rescaled AGP norm $||A_{\lambda}||^2/L$ for different models: A) Model: XXZ chain with NNN interaction (eqn. \ref{NNN_Ham}). a)  $\lambda=\Delta$ and b)  $\lambda=\Delta_2$. \textit{Parameters:} $ \Delta=1.1, \Delta_2=1$. B) Model: XXZ with defect in the middle (eqn.~\ref{Ham_XXZ_defect}). a)  $\lambda$ is chosen as $\Delta$. \textit{Inset:} AGP norm $||A_{\lambda}||^2$ for XXZ with defect in the middle model where  $\lambda$ is chosen as $\epsilon_d$.  \textit{Parameters:} $\epsilon_d=1, \Delta=1.1$. This is calculated for the full Hilbert space, not in any specific symmetry sector. }
\label{universal_slope}
\end{figure}

Here we study the AGP norm in the XXZ chain in the limit when the magnitude of the integrability-breaking perturbation (either the local defect energy $\epsilon_d$ or the NNN interaction strength $\Delta_2$) is of the same magnitude as the $\Delta/J$ energy scale. In this limit, we find that the AGP has an exponential scaling with system size characterized by an almost universal slope $\beta  \approx \log 2$, which is close to the one predicted by ETH. Details about the model and its parameters are given in the caption of Fig. \ref{universal_slope}.




\end{appendix}

\clearpage

\bibliography{ref_general} 

\end{document}